\documentclass[10pt]{article}
\usepackage{geometry} 
\usepackage[parfill]{parskip}    
\usepackage{graphicx} 
\usepackage{amssymb}
\usepackage{epstopdf}
\usepackage{alltt} 
\usepackage{rotating} 
\usepackage{rotfloat}
\usepackage{float} 
\usepackage{multirow}
\usepackage{fancyhdr} 
\usepackage{wrapfig}
\usepackage{amsmath}
\usepackage{subfigure}
\usepackage{fullpage}
\usepackage{hyperref}
\usepackage{overcite}
\usepackage{url}
\usepackage{textcomp}
\usepackage{color}
\usepackage[usernames, dvipnames]{xcolor}
\usepackage{gensymb}
\usepackage[utf8]{inputenc}
\usepackage[T1]{fontenc}
\usepackage{braket}
\usepackage{bm}
\usepackage{enumitem}
\usepackage{longtable}
\usepackage{booktabs}
\usepackage[flushleft]{threeparttable}

\usepackage{xcolor}

\usepackage[switch]{lineno}

\graphicspath{./}

\newcommand{\apj}{\it Astrophys. J.}
\newcommand{\aap}{Astron. Astrophys.}
\newcommand{\mnras}{\it Mon. Not. R. Astron. Soc.}
\newcommand{\nat}{\it Nat.}
\newcommand{\araa}{\it Annu. Rev. Astron. Astrophys.}

\newcommand{\apjl}{\it Astrophys. J. Letters.}
\newcommand{\aj}{\it Astron. J.}
\newcommand{\apjs}{\it Astron. J. Supp. S.}
\newcommand{\aaps}{\it Astron. Astrophys. Supp. S.}
\newcommand{\ssr}{Space Science Reviews}
\newcommand{\actaa}{Acta Astronomica}
\newcommand{\apss}{Astrophysics and Space Science}
\newcommand{\aapr}{The Astron. and Astrophys. Rev.}
\newcommand{\jcap}{Journal of Cosmology and Astroparticle Physics}

\begin{document}


\begin{center}
\LARGE{A calibration point for stellar evolution \\ from massive star asteroseismology}
\end{center}

\normalsize

\vspace{1cm}

\begin{center}	
Siemen~Burssens$^{1,\ast}$,
Dominic~M. Bowman$^{1, \ast}$,
Mathias~Michielsen$^{1}$,
Sergio~Sim{\'o}n-D{\' i}az$^{2,3}$,
Conny~Aerts$^{1,4,5}$, 
Vincent~Vanlaer$^{1}$,
Gareth~Banyard$^{1}$,
Nicolas~Nardetto$^{6}$, 
Richard~H.~D.~Townsend$^{7,8}$,
Gerald Handler$^{9}$,
Joey~S.~G.~Mombarg$^{10,1}$,
Roland~Vanderspek$^{11}$, 
George~Ricker$^{11}$.
\end{center}

\vspace{0.5cm}
	
$^{1}$ Institute of Astronomy, KU Leuven, Celestijnenlaan 200D, 3001 Leuven, Belgium,\\
$^{2}$ Instituto de Astrof{\' i}sica de Canarias, E-38200 La Laguna, Tenerife, Spain,\\
$^{3}$ Departamento de Astrof{\' i}sica, Universidad de La Laguna, E-38205 La Laguna, Tenerife, Spain,\\
$^{4}$ Department of Astrophysics, IMAPP, Radboud University Nijmegen,\\
NL-6500 GL Nijmegen, The Netherlands,\\
$^{5}$ Max Planck Institute for Astronomy, K\"{o}nigstuhl 17, 69117 Heidelberg, Germany,\\
$^{6}$ Universit\'e C\^ote d'Azur, Observatoire de la C\^ote d'Azur, CNRS, Laboratoire Lagrange, France,\\
$^{7}$ Department of Astronomy, University of Wisconsin-Madison, 2535 Sterling Hall,\\ 
475 N. Charter Street, Madison, WI 53706, USA,\\
$^{8}$ Kavli Institute for Theoretical Physics, University of California, Santa Barbara, CA 93106, USA,\\
$^{9}$ Nicolaus Copernicus Astronomical Center, Polish Academy of Sciences, Warsaw, Poland \\
$^{10}$ IRAP, Universit\'e de Toulouse, CNRS, UPS, CNES, 14 avenue \'Edouard Belin,\\ F-31400 Toulouse, France,\\
$^{11}$ Department of Physics and Kavli Institute for Astrophysics and Space Research,\\
Massachusetts Institute of Technology, Cambridge, MA 02139, USA 

\vfill
$^\ast$ Corresponding authors: siemen.burssens@gmail.com and dominic.bowman@kuleuven.be \\

\newpage

\section*{Abstract}

{\bf 
Massive stars are progenitors of supernovae, neutron stars and black holes. During the hydrogen-core burning phase 
their convective cores are the prime drivers of their evolution, but inferences of core masses are subject to unconstrained boundary mixing processes. Moreover, uncalibrated transport mechanisms can lead to strong envelope mixing and differential radial rotation. Ascertaining the efficiency of the transport mechanisms is challenging because of a lack of observational constraints. Here we deduce the convective core mass and robustly demonstrate non-rigid radial rotation in a supernova progenitor, the $12.0^{+1.5}_{-1.5}$ solar-mass hydrogen-burning star HD~192575, using asteroseismology, TESS photometry, high-resolution spectroscopy, and Gaia astrometry. We infer a convective core mass ($M_{\rm cc} = 2.9^{+0.5}_{-0.8}$ solar masses), and find the core to be rotating between 1.4 and 6.3 times faster than the stellar envelope depending on the location of the rotational shear layer. Our results deliver a robust inferred core mass of a massive star using asteroseismology from space-based photometry. HD~192575 is a unique anchor point for studying interior rotation and mixing processes, and thus also angular momentum transport mechanisms inside massive stars.
}

\vspace{0.5cm}
\section*{Main text}
\setcounter{section}{1}
\setcounter{subsection}{0}
\subsection*{Chemical mixing and angular momentum transport}
The physical processes that set the convective core mass and interior rotation profile of a massive star are among the largest uncertainties in stellar evolution theory. They are a result of the strong influence of the star formation mechanism, the possible interaction or merger with a companion star or the potential presence of a magnetic field, all of which modify the interior mixing and rotation profiles and the star's subsequent evolution \cite{Maeder2009,Langer2012}. As a consequence of a historic lack of observations that probe stellar interiors, the mechanisms transporting chemical species and angular momentum in massive stars are essentially uncalibrated. In turn, models with varying prescriptions for such processes predict significantly different chemical yields and spin rates of compact objects after a supernova explosion.

A particularly successful technique and modelling strategy to unlock the hidden interiors of stars is asteroseismology \cite{Aerts2021}, which uses the eigenfrequencies of a star identified from a time series of its observed surface properties to determine its structure. The NASA Kepler mission \cite{Borucki2010} revealed the interior rotation profiles of stars with masses ranging between $0.8$ and $8$~M$_{\odot}$, from the main sequence through to the red giant phase and beyond \cite{Gehan2018,Pedersen2021}. The approximately rigid radial rotation profiles of low-mass main-sequence stars, together with the core-to-envelope rotation ratios between approximately 1 and 10 for their evolved counterparts revealed large discrepancies in the theory of angular momentum transport across stellar evolution \cite{Aerts2019a}. However, similar inferences for stars with birth masses above $8$~M$_{\odot}$ are lacking due to the insufficient time-series observations of such stars \cite{Bowman2020b}. This includes the $\beta$~Cep stars ($8$ M$_{\odot}$ $\lesssim$ M$_{\star} \lesssim 30$~M$_{\odot}$), which are found between the zero-age main sequence (ZAMS) through to beyond the terminal-age main sequence (TAMS), and have spectral types between late-O and early-B \cite{Stankov2005}. The $\beta$~Cep stars exhibit gravity and pressure modes that directly probe the deep interior and envelope physical conditions, respectively\cite{Aerts2003b,Dupret2004,Pamyatnykh2004,Dziembowski2008}.

The NASA TESS (Transiting Exoplanet Survey Satellite) space mission successfully delivers high-precision light curves over almost the full sky since July 2018\cite{Ricker2015}. This phenomenally increased the number of massive stars with high-precision long-term light curves. These TESS data led to our discovery of the $\beta$~Cep star HD~192575 with unprecedented pulsation mode frequencies for asteroseismic modelling, allowing for the determination of robust estimates of its age and core mass, and constrain its differential radial rotation profile. Thus, HD~192575 stands as a unique anchor point for the theory of stellar evolution of massive stars.

\subsection*{Observations with TESS, HERMES and Gaia}

\begin{figure}
    \centering
	\includegraphics[scale = 0.70]{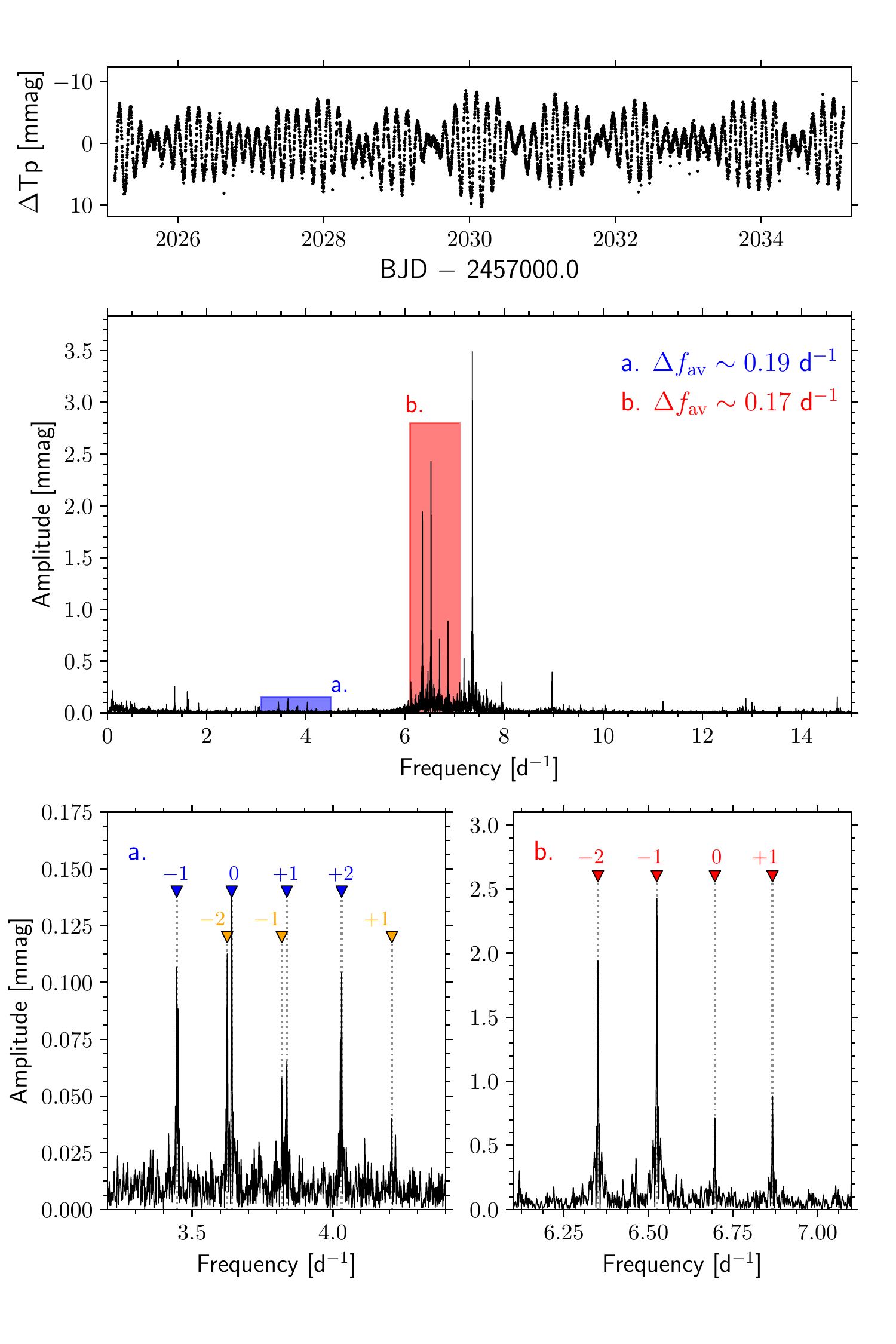}
    \caption{\textbf{TESS data of the $\beta$~Cep star HD~192575.} The top panel shows a $10$~d cut-out of the TESS light curve, with differential flux in the TESS wavelength passband ($\Delta T_{\rm p}$) with units of milli-magnitude (mmag) versus time in units of Barycentric Julian Date (BJD). The middle panel shows the LS periodogram of the 324-d TESS light curve. Two important frequency ranges (shown in panels a and b) encompassing three important rotational multiplets are highlighted, each having a different average frequency splittings. The two lower panels show the multiplets and label the identified azimuthal order in greater detail, with the measured frequencies in each multiplet shown by triangles with colours corresponding to the multiplets in the full periodogram. The lower-frequency multiplet shows two sets of overlapping modes, which is caused by an avoided crossing.}
\end{figure}

Prior to this work, HD~192575 had received little attention (see Methods). We utilised the 2-min cadence cycle~2 TESS data, which have a time base of $\Delta \rm T=323.77$~d and a duty cycle of $75.7\%$. Calculation of a Lomb-Scargle (LS) periodogram of these TESS data revealed several dominant pulsations, many
of which belong to rotational multiplets (see Figure~1). Pulsation modes are characterised by their angular degree $\ell$, azimuthal order $m$, and radial order $n_{\rm pg}$. In a spherically symmetric star, frequencies of the same $\ell$ and $n_{\rm pg}$ but different $m$ are degenerate. Rotation effectively lifts this degeneracy, perturbing mode frequencies with different $m$ values \cite{Unno1989} and thus allowing for mode geometry identification, which is an important prerequisite for asteroseismic modelling.

We extracted all the significant frequencies in the cycle~2 TESS light curve of HD~192575 and deduced the frequency separations of three confidently identified multiplets and two candidate multiplets (see Methods). We used the three confidently identified multiplets to perform mode identification and measured the separation of frequencies within each rotational multiplet\cite{Aerts2003b} to infer the rotation rate within the pulsation cavity of the involved modes (see Methods). These include two overlapping multiplets that are explained as two $\ell=2$ modes approaching or just passing an avoided crossing. This phenomenon occurs when g and p modes of the same angular degree interact at specific evolutionary stages due to a star's changing structure as a function of evolution\cite{Aizenman1977,Unno1989,Mazumdar2006}. 

We determined the atmospheric parameters of HD\,192575 from four high-resolution optical spectra with the HERMES spectrograph ($R\approx85000$) mounted on the 1.2-m Mercator Telescope in La Palma, Spain \cite{Raskin2011}. We found no evidence for short- or intermediate-period binarity (see Methods). We derived a projected surface rotation velocity of $v\,\sin\,i =27^{+6}_{-8}$~km~s$^{-1}$, an effective temperature of $T_{\rm eff} = 23900\pm 900$~K, and a surface gravity of $\log g = 3.65\pm0.15$~dex, which place
HD~192575 within the $\beta$~Cep instability region\cite{Walczak2015} (see Methods). Using the Gaia-DR3 parallax, we derived a luminosity of $\log L_{\star}/{\rm L_{\odot}}=4.30\pm0.07$ (see Methods). The location of HD~192575 in the Hertzsprung--Russell diagram allowed us to derive the $n_{\rm pg}$ values of three rotational multiplets
unambiguously, thus yielding firm mode identifications for three multiplets (see Methods). This includes two gravity mode multiplets, with $(\ell, n_{\rm pg})= (2,-2)$ and $(2,-1)$ and one pressure mode multiplet $(\ell, n_{\rm pg})= (2,2)$. For two additional multiplets we derived both $\ell$ and $n_{\rm pg}$ but the $m$ values remained less certain.

\subsection*{A robust convective core mass of HD~192575}

\begin{figure}
    \centering
	\includegraphics[scale = 0.5]{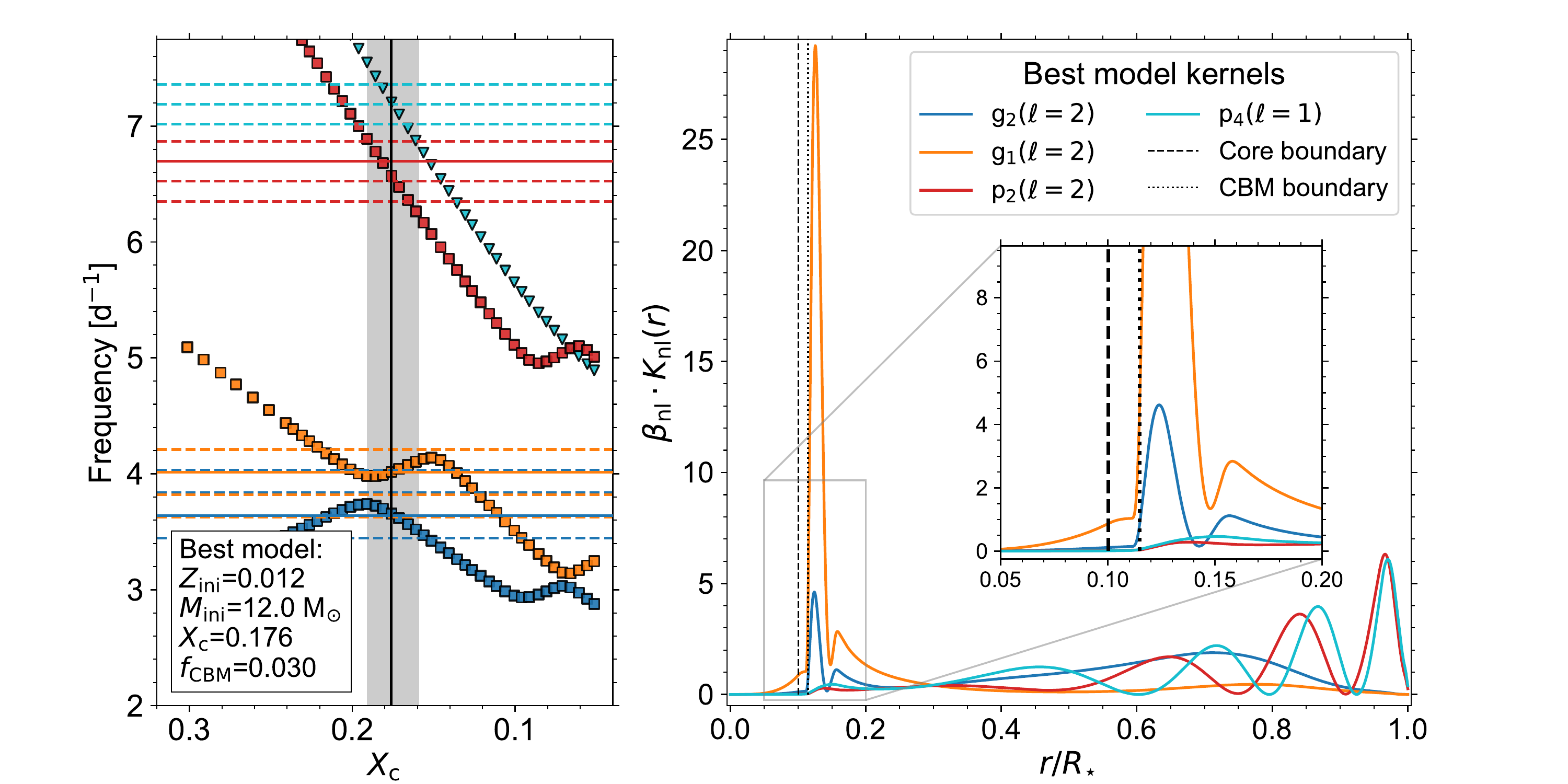}
    \caption{\textbf{Dominant pulsation modes in the best structure model of HD~192575 and their rotational kernels}. The left panel shows the evolution of the theoretical frequencies of the best evolution model for the confidently identified multiplets and the multiplet containing the dominant frequency ($f_{1}$) as a function of core hydrogen content ($X_{\rm c}$) as a proxy of stellar age. The best structure model is marked by a black vertical line. The $X_c$ values allowed by the avoided crossing criterion and by the $2\sigma$~$T_{\rm eff}- \log L/{\rm L_{\odot}}$ confidence intervals for the best model for the evolution track are shown by the grey area. The fitted and non-fitted component frequencies of the observed rotational multiplets are given as solid and dashed horizontal lines, respectively. The frequency precision of the individual frequencies is of the order of the line width. The right panel shows the rotational kernels, $\beta_{\rm nl}K_{\rm nl} (r)$, for the confidently identified multiplets in the left panel as a function of the normalised stellar radius, including pressure and gravity modes marked by p and g with the subscript referring to their $|n_{\rm pg}|$ value. We also show in red the kernel for the multiplet containing the dominant frequency, $f_{1}$, for which we derived $(\ell, n_{\rm pg})$ but could not confidently identify $m$. The vertical dashed and dotted lines show the core boundary and edge of the convective boundary mixing (CBM) region, respectively. The colours of the left and right panels refer to the same pulsation mode geometries.}
\end{figure}

Using the confidently identified rotational multiplets (having $\ell,n_{\rm pg}$ and $m$) and the $2\sigma$ confidence interval for the spectroscopic $T_{\rm eff}$ and Gaia-informed $\log L_{\star}/{\rm L_{\odot}}$ values, we performed a comparison of observed pulsation mode frequencies with those calculated from a dense grid of
stellar structure models (see Methods). We used the Mahalanobis distance (MD) as the merit function to be minimised, given as:
\begin{equation}\label{eq:MD}
    \theta_{0} =\text{arg}\underset{\bm \theta}{\text{min}} \{(\bm{Y^{\rm theo}(\theta)} - \bm{Y^{\rm obs}})^\top (V+\Lambda)^{-1} (\bm{Y^{\rm theo}(\theta)} -\bm{Y^{\rm obs}})\},
\end{equation}
where $\bm{\theta}=(\theta^{1}, \theta^{2},...,\theta^{n})$ are the $n$ parameters to be estimated (e.g. mass and age), $\bm{Y^{\rm theo}(\theta)}$ and $\bm{Y^{\rm obs}}$ are the theoretical predictions and observables (in our case the pulsation frequencies), respectively, $V=\text{var}(\bm{Y^{\rm theo}})$ is the variance-covariance matrix including the correlation structure within the theoretical grid, and $\Lambda$ is the matrix with the observed pulsation frequency uncertainties. The use of Eq.~(\ref{eq:MD}) is more sophisticated than a $\chi^{2}$ merit function, because it includes the covariance matrices $V$ and $\Lambda$\cite{Aerts2018b}. This allows for statistically appropriate parameter estimation, including systematic uncertainties due to limited parameterisations of physical prescriptions within models, such as convective boundary mixing\cite{Michielsen2021}. 

We constrained a minimal set of free parameters $\bm \theta = (Z_{\rm ini}, M_{\rm ini}, X_{\rm c}, f_{\rm CBM})$ which are the initial metallicity, mass, core hydrogen content, and a parameterisation of the convective boundary mixing. Our large grid of 39312 structure models was calculated using the stellar evolution code MESA\cite{Paxton2019} and corresponding pulsation mode frequencies using the stellar oscillation code GYRE\cite{Townsend2018} (see Methods). We filter models that do not satisfy the observed avoided crossing between the modes in the overlapping multiplet (see Figures~1,~2). That is, we require that the theoretical frequencies of a structure model for the overlapping $(\ell, n_{\rm pg}, m)= (2,-2, 0)$ and $ (2,-1, 0)$ modes of the avoided crossing, $f_{\rm theo, g2}$ and $f_{\rm theo, g1}$, to be maximally separated by: 
\begin{equation}\label{eq:avoided_crossing}
   f_{\rm theo, g1} - f_{\rm theo, g2} \leq (f_{\rm obs, g1} - f_{\rm obs, g2}) + \sigma f_{\rm obs, g1} + \sigma f_{\rm obs, g2},
\end{equation}
where $f_{\rm obs, g1}, f_{\rm obs, g2} $, $\sigma f_{\rm obs, g1} $ and $\sigma f_{\rm obs, g2} $ are the measured frequencies and uncertainties of the identified $(\ell, n_{\rm pg}, m)= (2,-2, 0)$ and $ (2,-1, 0)$ modes undergoing the avoided crossing (see Methods). The uncertainties are of the order $10^{-5}$ -- $10^{-6}$~d$^{-1}$. The avoided crossing requirement allows for a maximum separation of $\sim 0.37$~d$^{-1}$. Enforcing this requirement of the avoided crossing of the identified rotational multiplets allowed us to calculate the variance-covariance matrix of the remaining 4962 acceptable structure models. We calculated the MD value for these structure models using the theoretical and observed $m=0$ frequencies of the three confidently identified multiplets and then considered only those that satisfied the $2\sigma$ confidence intervals in the HR~diagram (i.e. $T_{\rm eff}$ and $\log L_{\star}/ {\rm L_{\odot}}$). This yielded 663 models that satisfied both the requirement of the avoided crossing and the $2\sigma$ confidence intervals in the HR~diagram (see left panel of Figure~2 for an illustration in case of the best model stellar evolution track). We show the parameter correlation diagram of the 663 models for the varied set of parameters $\bm \theta$ in Figure~3. We subsequently determined the uncertainties for the three free parameters for the best-fitting model using the likelihood function associated with the Mahalanobis distance (Eq.~\ref{eq:MD}) and Bayes' theorem (see Methods).

\begin{figure}
    \centering
	\includegraphics[scale = 0.5]{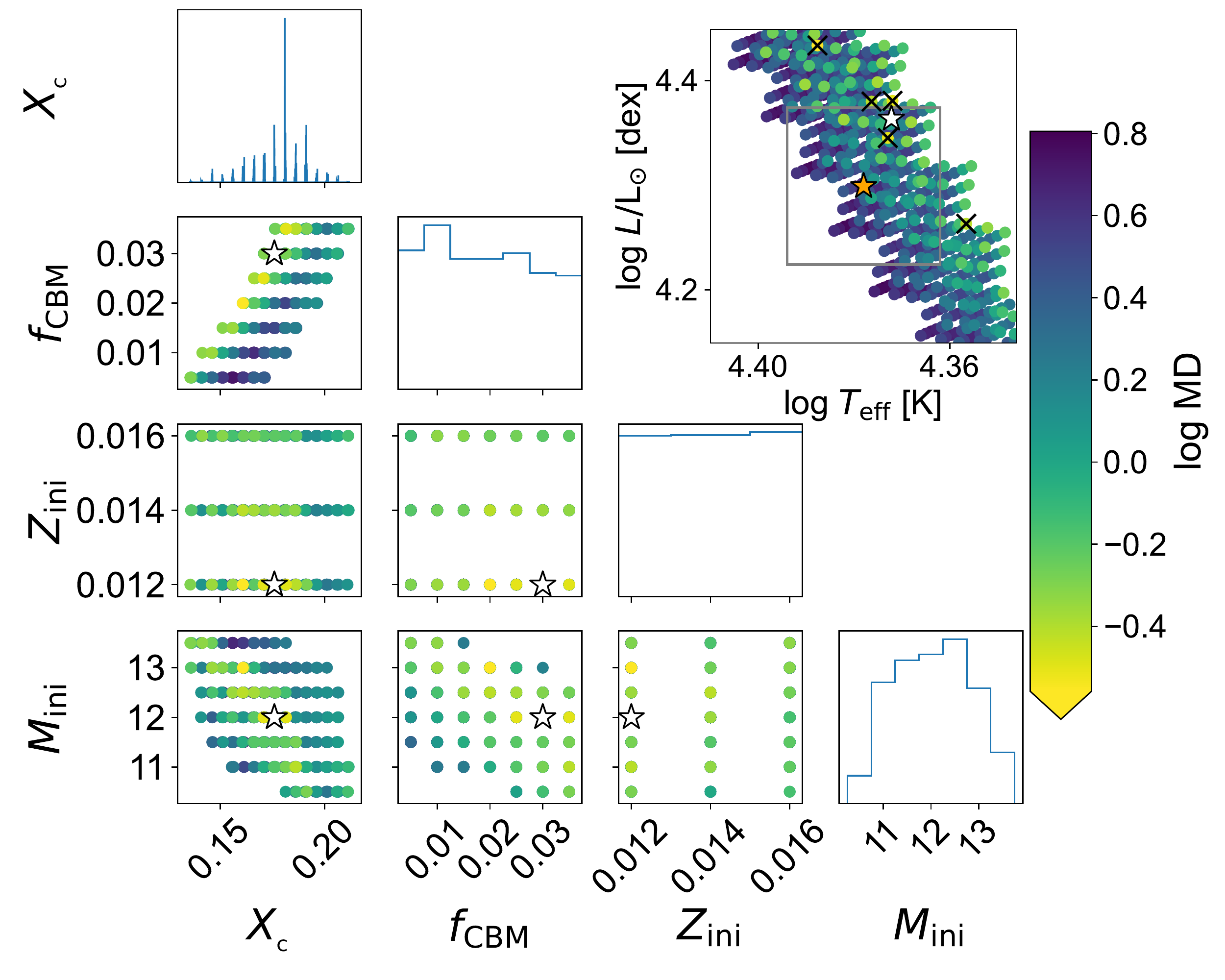}
    \caption{\textbf{Forward asteroseismic modelling results overview.} The panels below the diagonal show parameter correlations of the best models using the MD. The panels on the diagonal show binned parameter distributions of the parameter marked on the horizontal axis. The overall best fitting model is shown as a white star, with the black crosses marking the next five best models in the HR~diagram in the top-right panel. The orange star and grey box indicate the spectroscopic parameters and 
    the $1\sigma$ ($\log T_{\rm eff}-\log L_{\star}/{\rm L_{\odot}}$) confidence intervals of the star, respectively. Only models in the $2\sigma$ ($\log T_{\rm eff}-\log L_{\star}/{\rm L_{\odot}}$) confidence intervals and that satisfy the observed avoided crossing of the $(\ell, n_{\rm pg}, m)= (2,-2, 0)$ and $ (2,-1, 0)$ modes are shown, colour-coded by their MD value.}
\end{figure}

This work provides robust constraints on the interior properties of a supernova progenitor using asteroseismology: 
HD~192575 has a mass of $M_{\star} = 12.0^{+1.5}_{-1.5}$~M$_{\odot}$, a core hydrogen mass fraction of $X_{\rm c}= 0.176^{+0.035}_{-0.040}$, a convective boundary mixing value of $f_{\rm CBM}=0.030^{+0.005}_{-0.025}$, and we infer an age of $t_{\star}=17.0^{+4.7}_{-5.4}$~Myr, and a radius of $R_{\star, \rm seism}=9.1^{+0.8}_{-1.7}$~R$_{\odot}$ ($2\sigma$ confidence intervals). Using the models that fall within the 2$\sigma$ confidence interval of the best model and Bayes' theorem, we infer a core mass of $M_{\rm cc}= 2.9^{+0.5}_{-0.8}$~M$_{\odot}$ and core radius $R_{\rm cc}= 0.91^{+0.11}_{-0.15}$~R$_{\odot}$(see Methods). We derived an initial metallicity of $Z_{\rm ini}=0.012^{+0.004}_{-0.000}$ at the lower end of our considered range suggesting our methodology is not sensitive to such a parameter when using only three identified pulsation frequencies, but this is to be expected. We nevertheless include it as a varied parameter due to the known mass-metallicity-overshooting correlation\cite{Ausseloos2004}, and to accurately describe the central chemically homogeneous regions\cite{Salmon2022a}. Indeed, by including $Z_{\rm ini}$ as a varied parameter we find a lower initial mass and metallicity, but a higher CBM value. Whilst the confidence intervals on other parameters remained the same, the derived confidence intervals on the core $R_{\rm cc}$ and stellar radii $R_{\star, \rm seism}$ increased slightly compared to the fixed initial metallicity case. This is expected since chemical composition affects the core boundary, the $\mu$-gradient zone, and the stellar atmosphere through the opacity. As a result, our methodology yields an accurate convective core mass of a massive star that incorporates a multitude of theoretical uncertainties within stellar structure models of massive pulsating stars. These constraints on the fundamental parameters of HD~192575 from asteroseismology have fractional uncertainties of order 10-25\%, which is a level of precision unparalleled for single supernova progenitors (see Methods). Given the advanced stage of core hydrogen burning of HD~192575, its core mass and radius have already decreased dramatically since birth. Indeed, for the best stellar structure model, the core mass and radius at the ZAMS are $M_{\rm cc, ZAMS}=4.4$~M$_{\odot}$ and $R_{\rm cc, ZAMS}=1.1$~R$_{\odot}$, respectively. Our result is the first inference of the core mass in a single massive star using modern asteroseismic techniques that appropriately takes a variety of systematic theoretical uncertainties and parameter correlations into account.

\subsection*{Differential radial rotation in HD~192575 }\label{main:diff_rot}

\begin{figure}
    \centering
	\includegraphics[scale = 0.75]{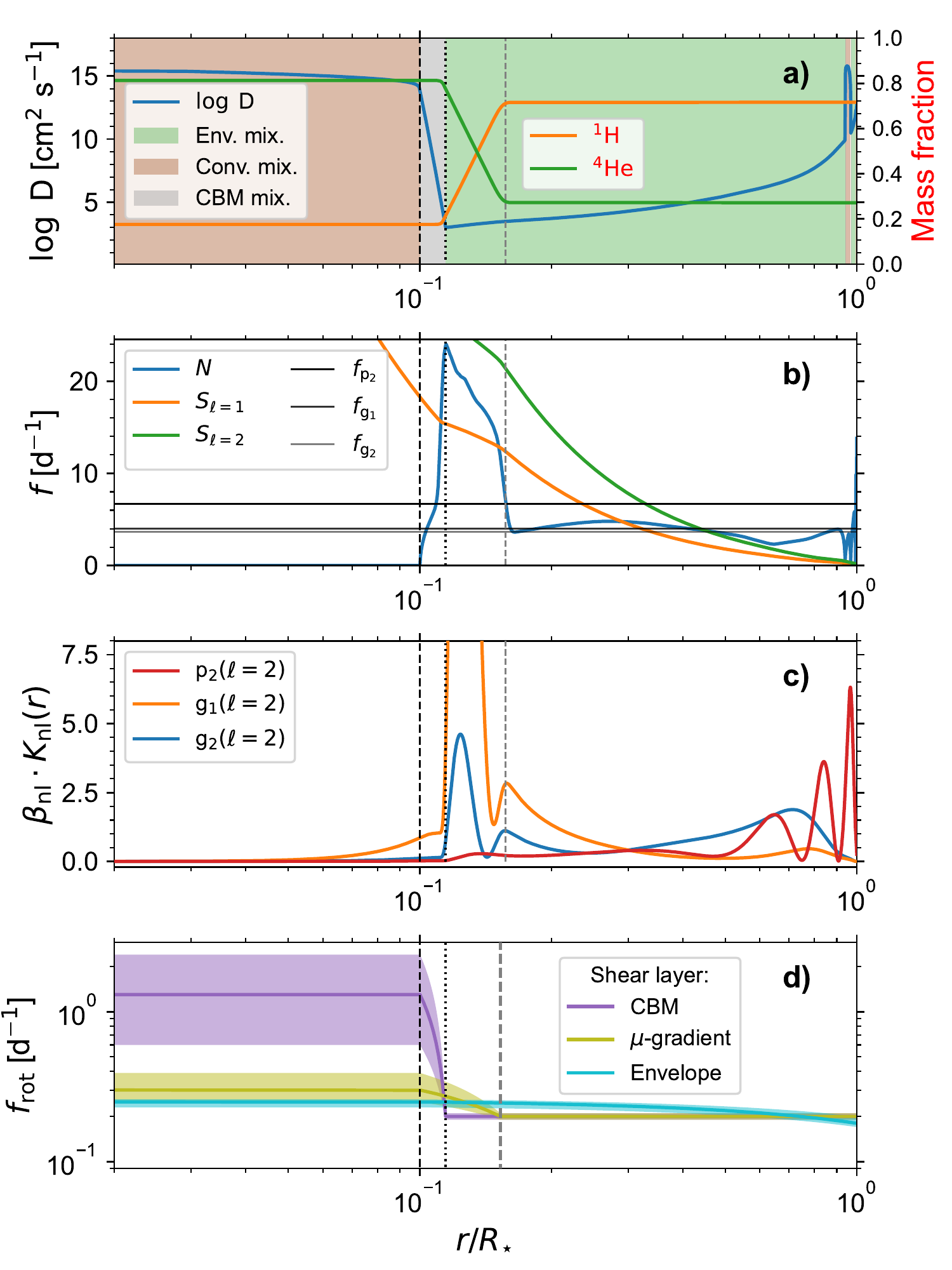}
    \caption{\textbf{Structure of the best-fitting model of HD~192575.} Panel a) shows the mixing properties of the model, with the blue line indicating the level of mixing, the shaded regions represent different mixing regimes, and the orange and green lines denote the fractional mass of $^1\rm H$ and $^4\rm He$ respectively (the values of the latter two parameters are given by the axis on the right). Panel b) shows the propagation diagram, including the Brunt-V\"{a}is\"{a}l\"{a} frequency, $N$, and Lamb frequencies, $S_{\ell}$, for $\ell=1,2$. The values of the zonal mode frequencies of the considered rotational multiplets are shown as horizontal lines. Panel c) shows a zoom-in of the rotational kernels $\beta_{\rm nl}K_{\rm nl}$ for the identified zonal modes used to derive the rotation profile.  Panel d) shows the three-zone rotation profiles throughout the stellar interior for our three different choices for the rotational shear layer: the convective boundary mixing (CBM) layer, the $\mu$-gradient layer, or the entire envelope. The black vertical dashed and dotted lines on each panel indicate the transition region between the convective core and the radiative envelope. For reference, the position of the convective core at the ZAMS is shown by a grey dashed line.}
\end{figure}

In addition to fundamental parameters, we deduce the radial rotation profile of HD\,192575 from its accurate stellar structure model under three different assumptions on the radial rotation profile shape. The star's observed rotational splittings and its low radial order mode frequencies, along with the high frequency precision, allow the use of a first-order perturbative approach for the Coriolis acceleration. In that approximation, pulsation mode frequencies are given by:
\begin{equation}\label{eq:split_slow_rot}
    f_{nlm} = f_{nl0} + m \beta_{nl} \int_{0}^{R} K_{nl} f_{\rm rot}(r) dr, 
\end{equation}
where $K_{nl}$ is the rotational kernel, $f_{\rm rot}(r)$ the cyclic rotation frequency profile, and $\beta_{nl} = (1-C_{nl})$ with $C_{nl}$ being the Ledoux constant \cite{Ledoux1951}. Both $\beta_{nl}$ and $K_{nl}$ depend on the mode identification and the structure of the
star\cite{Ledoux1951}. The different frequency separations within the detected rotational multiplets allow us to rule out rigid rotation for HD~192575 (see Methods). We show the rotational kernels for identified modes of the best model in the right panel of Figure~2, demonstrating the strong probing power of the near-core region of HD~192575's low radial order gravity modes. 

The structure of the near-core region of the best model is shown in Figure~4. This reveals efficient trapping of pulsation modes in the $\mu$-gradient zone left behind by the receding convective core. The differences in the probing power of the identified rotational multiplets provide direct inference of the core, near-core and envelope rotation rate.

To quantify the differential radial rotation profile of HD~192575 we use an analytical three-zone model of the rotation rate $f_{\rm rot}(x)$ as a function of the fractional radius $x$:
\begin{equation}\label{eq:omega_model}
  f_{\rm rot} (x) =
    \begin{cases}
      f_{\rm cc}& \text{if $x\leq x_{\rm cc}$}\\
      f_{\rm cc} + (f_{\rm cc} -f_{\rm env}) \frac{x-x_{\rm cc}}{x_{\rm cc}-x_{\rm shear}}& \text{if $x_{\rm cc} \leq x \leq x_{\rm shear}$}\\
      f_{\rm env} & \text{if $x\geq x_{\rm shear}$},
    \end{cases}       
\end{equation}
where $f_{\rm cc}, f_{\rm env}$ are uniform rotation frequencies in the core and envelope respectively, and $x_{\rm cc}, x_{\rm shear}$ the fractional radii of the core and rotational shear layer, respectively \cite{Aerts2003b,Pamyatnykh2004,Dziembowski2008}. This three-zone model constitutes a rotating core and a rotating envelope, separated by a
transition shear layer in which we assume the rotation rate to change linearly. It is currently unknown where such a shear layer should be situated in massive stars, hence we are inspired by the few examples in the literature\cite{Aerts2003b, Pamyatnykh2004, Dupret2004} to test the sensitivity of different assumed scenarios in our study. Furthermore, it has been shown in previous work on lower-mass stars that the rotation profile that reproduces the measured frequency separations in rotational multiplets strongly depends on the assumed shear layer (and shape)\cite{Deheuvels2012, Deheuvels2014}. Rotation-induced shear in main-sequence massive stars could mainly occur in the narrow convective boundary mixing layer (CBM) or in the larger $\mu$-gradient zone or even in the entire envelope (see Figure~4). Therefore, in this work, we test these three different formulations for the position and size of the shear layer, estimating the free parameters $f_{\rm cc}$ and $f_{\rm env}$ from the measured frequency splittings by constructing a system of two equations using Eq.~(\ref{eq:split_slow_rot}) and Eq.~(\ref{eq:omega_model}) (see Methods).

Assuming that the shear layer is the CBM region, we propagate the uncertainties in the best-fitting structure model forward into the calculation of the differential radial rotation profile for all models contained within its 2$\sigma$ confidence interval, and derive $f_{\rm cc}/f_{\rm env} = 6.3^{+6.5}_{-3.8}$ with $f_{\rm cc} = 1.3^{+1.1}_{-0.7}$~d$^{-1}$ and $f_{\rm env} = 0.20^{+0.01}_{-0.01}$~d$^{-1}$ for the core and envelope rotation frequencies, respectively (see Methods). 
On the other hand, when assuming that the shear layer coincides with the $\mu$-gradient zone, we derive $f_{\rm cc}/f_{\rm env} =1.49^{+0.56}_{-0.33}$ with $f_{\rm cc} = 0.30^{+0.09}_{-0.05}$~d$^{-1}$ and $f_{\rm env} = 0.20^{+0.01}_{-0.01}$~d$^{-1}$ for the core and envelope rotation frequencies, respectively. Finally, when assuming that rotational shear occurs linearly in the whole radiative envelope, we derive $f_{\rm cc}/f_{\rm env} =1.40^{+0.16}_{-0.20}$ with $f_{\rm cc} =  0.25^{+0.01}_{-0.02}$~d$^{-1}$ and $f_{\rm env} = 0.18^{+0.02}_{-0.01}$~d$^{-1}$ for the core and envelope rotation frequencies, respectively (see Methods). We summarise these values in Table~1 and show confidence intervals for the rotation profiles in Figure~4. For all three assumptions on the position and size of the shear layer, we find strong evidence for a non-rigid radial rotation profile in HD\,192575 with the core rotating 1.4 to 6.3 times faster than the envelope. Using these results we further infer that HD~192575 is observed at inclination $i=17(6)\degree$ in case of a CBM or a $\mu$-gradient zone shear layer and $i=19(6)\degree$ in case of shear occurring linearly in the whole radiative envelope.

\subsection*{Discussion}

Differential rotation profiles have been studied in only a few $\beta$~Cep stars\cite{Aerts2003b, Pamyatnykh2004, Dupret2004}. We show the main-sequence massive stars with measured $f_{\rm cc}/f_{\rm env}$ as a function of $T_{\rm eff}$ in Figure~5. Due to limitations in the available data and computing power at the time these studies were restricted to using only a handful of stellar structure models. For similar reasons all of these studies also assumed only a single location of the rotational shear layer (except one, which used structure models with shellular rotation) and could not test the sensitivity of their inferred rotation profile to this assumption. It is because of the long time base and unparalleled frequency precision of the cycle`2 TESS data of HD~192575 that we can confirm such rotation profiles are strongly sensitive to this assumption. Indeed, depending on the assumption for the rotational shear layer inside HD~192575, the star is either the most or the least differentially rotating single $\beta$~Cep star (see Table~1). This makes HD~192575 a massive star with tight constraints on mass, age, radius and convective core mass taking proper account of correlations among the parameters and a multitude of systematic theoretical uncertainties for their estimation, in addition to robust core-to-surface rotation profiles assuming three different shear layers. 

Standard one-dimensional rotating massive star structure models typically include meridional circulation and horizontal turbulence as angular momentum transport mechanisms\cite{Georgy2013a}. These models predict differentiality at the level of $f_{\rm cc}/f_{\rm env}\approx 2-4$ for 10.5-13.5~M$_{\odot}$ stars approaching the TAMS ($X_{\rm c}<0.20$) with $v_{\rm eq, ini}/v_{\rm crit}= 10-50$~\% and which started their lives with uniform rotation at the ZAMS. These specific values were retrieved from stellar models made available by Georgy et al. 2013\cite{Georgy2013a} (\url{https://www.unige.ch/sciences/astro/evolution/fr/base-de-donnees/syclist/}). A few stellar structure model simulations also include magnetic instabilities, such as the Taylor-Spruit(-Fuller) dynamo\cite{Spruit2002, Fuller2019}, which are triggered by differential rotation and efficiently redistribute angular momentum. These stellar structure models generally result in lower differentiality than the models including standard hydrodynamical angular momentum transport\cite{Fuller2019, Salmon2022b}. 

Our analysis of HD~192575 when assuming a CBM-based shear layer thus reveals a two times larger value for $f_{\rm cc}/f_{\rm env}$ than angular momentum transport mechanisms (with or without magnetic instabilities) predict, as shown in Figure~5. This either implies that angular momentum transport in HD 192575 is more efficient than what is predicted from the standard transport mechanisms, that there is stronger removal of angular momentum at the surface by stellar winds than what is predicted from stellar wind models, or that the radial rotation profile of HD~192575 was already non-rigid at the ZAMS. The latter is supported by recent work on backtracing the present-day rotation profile of the $\beta$~Cep star HD~129929\cite{Aerts2003b, Dupret2004} to find non-rigid rotation ($\sim 1.74$) at the ZAMS\cite{Salmon2022b} (see Table~1). Non-rigid rotation at the ZAMS is supported by stars arriving at the ZAMS with high angular momentum\cite{Granada2014}. The aforementioned work on HD~129929 further finds that stellar structure models including magnetic instabilities underpredict the measured differentiality, but that hydrodynamic processes alone provide good agreement. On the other hand, when the $\mu$-gradient zone is assumed to be the shear layer, as was previously assumed for the $\beta\,$Cep stars $\nu\,$Eri\cite{Pamyatnykh2004,Dziembowski2008,Suarez2009} and 12~Lac\cite{Dziembowski2008}, we find lower differential rotation. This can be explained in two ways. Either HD~192575 had a very high initial rotation rate at the ZAMS ($v_{\rm eq, ini}/v_{\rm crit}\geq 50$~\%), or stronger angular momentum transport occurs in HD~192575 than predicted from stellar structure models which do not include magnetic instabilities\cite{Georgy2013a}. The limits on the differential rotation of HD~192575 from asteroseismology therefore provide important anchor points for future theoretical studies of the shear layer position and the angular momentum transport in massive stars\cite{Salmon2022b}.

HD~192575 will undergo a core-collapse supernova and leave behind a neutron star. The $f_{\rm cc}/f_{\rm env}$ values derived from different rotation profile shapes provide much-needed context to improve theoretical predictions of angular momentum transport in massive stars and neutron star spins, which currently show discrepancies with observations\cite{Langer2012}. 
Future avenues include moving from the analytical profile shapes assumed here, to more physical profile shapes informed by multi-dimensional stellar structure simulations such as those provided by the ESTER code\cite{Rieutord2016}. Additionally, to compare the results of HD~192575 with other $\beta$~Cep stars in Table~1, ensemble modelling of dozens of stars is a logical next step, by using similar treatments of rotation, parameter grids exploring a variety of theoretical formalisms, and statistical approaches to parameter estimation. Furthermore, motivated by the nearly uniform rotational splitting of the observed modes, we considered first-order perturbative effects to treat rotation (and stellar structure) in this study of HD~192575. Nevertheless, the relatively small (yet resolved) measured asymmetries will be used in future work to further constrain the interior rotation profile of HD~192575 using theoretical frameworks that include higher-order effects of stellar rotation in the mode calculations\cite{Suarez2009} and/or inversion analyses\cite{Triana2015}. Indeed, future rotation and structure inversions from all detected multiplets and the best-fitting structure models determined in this work will allow us to improve the characterisation of the interior physics. This includes the rotation profile but also, for example, the magnitude and shape of the CBM and envelope mixing, even further. Similarly, additional insight from the detected avoided crossing into the interior physics of HD~192575 may be gained by adapting frameworks developed in work on lower-mass stars\cite{Deheuvels2011}. Finally, the methodology developed in this work holds the potential to deliver the rotation profiles of many more pulsating massive stars whose TESS or future PLATO\cite{Rauer2014} data facilitate asteroseismology.

\begin{table}
\centering
\caption{\textbf{Asteroseismic estimates of $f_{\rm cc}/f_{\rm env}$ in HD~192575 and the massive star regime.} We note that for the majority of these literature values, there is a large amount of uncertainty and usually only an upper or lower limit for the core-envelope rotation ratio being available. For example, in the case of V836~Cen the confidence interval by Aerts et al. 2003\cite{Aerts2003b} resulted from a linear fit to one structure model, while for $\nu$~Eri the confidence interval resulted from the estimation of the profile in six arbitrarily chosen structure models to demonstrate the effect of different opacity tables, metal mixtures, and convective boundary mixing. For $\nu$~Eri we also include the work by Suarez et al. 2009\cite{Suarez2009} who used stellar structure and pulsation models with shellular rotation to derive the rotation profile.}
\label{table:ang_mom}
\renewcommand{\arraystretch}{1.2}
\begin{tabular}{c c c c c} 
 \hline
Star & $\log\,T_{\rm eff}$ & $f_{\rm cc}/f_{\rm env}$ & Assumed shear layer & Reference \\
& [K]  & && \\
\hline

V836~Cen (HD~129929) & $4.365\pm 0.015$ & $3.2\pm0.1$, $3.6$, $3.84$& Envelope & \cite{Aerts2003b, Dupret2004, Salmon2022b}\\
$\theta$~Oph (HD~157056) & $4.360\pm 0.018$ & 0.94-1.00 & Envelope & \cite{Briquet2007,Walczak2015}\\
12~Lac (HD~214993) & $4.375\pm 0.018$ & 1.00-4.65 &$\mu$-gradient &\cite{Handler2006, Dziembowski2008}\\
$\nu$~Eri (HD~29248) & $4.346\pm 0.011$ & 
$\left\{\begin{matrix}
 5.59\pm0.23  \\
2.58-2.64
\end{matrix}\right.$ & 
$\begin{matrix}
\text{$\mu$-gradient} \\
\text{Shellular}
\end{matrix} $&
$\begin{matrix}
\text{\cite{Pamyatnykh2004, Dziembowski2008}}\\
\text{\cite{Suarez2009}}
\end{matrix}$\\

HD~192575 & $4.378\pm 0.016$ & 
$\left\{\begin{matrix}
6.3^{+6.5}_{-3.8} \\
1.49^{+0.56}_{-0.33}\\ 
1.40^{+0.16}_{-0.20}
\end{matrix}\right.$ & 
$\begin{matrix}
\text{CBM}\\
\text{$\mu$-gradient} \\
\text{Envelope}
\end{matrix} $&
$\begin{matrix}
\text{This work}\\
\text{This work}\\
\text{This work}
\end{matrix}$\\

\hline

\end{tabular}

\end{table}

\renewcommand{\arraystretch}{1}
\begin{figure}
    \centering
	\includegraphics[scale = 0.60]{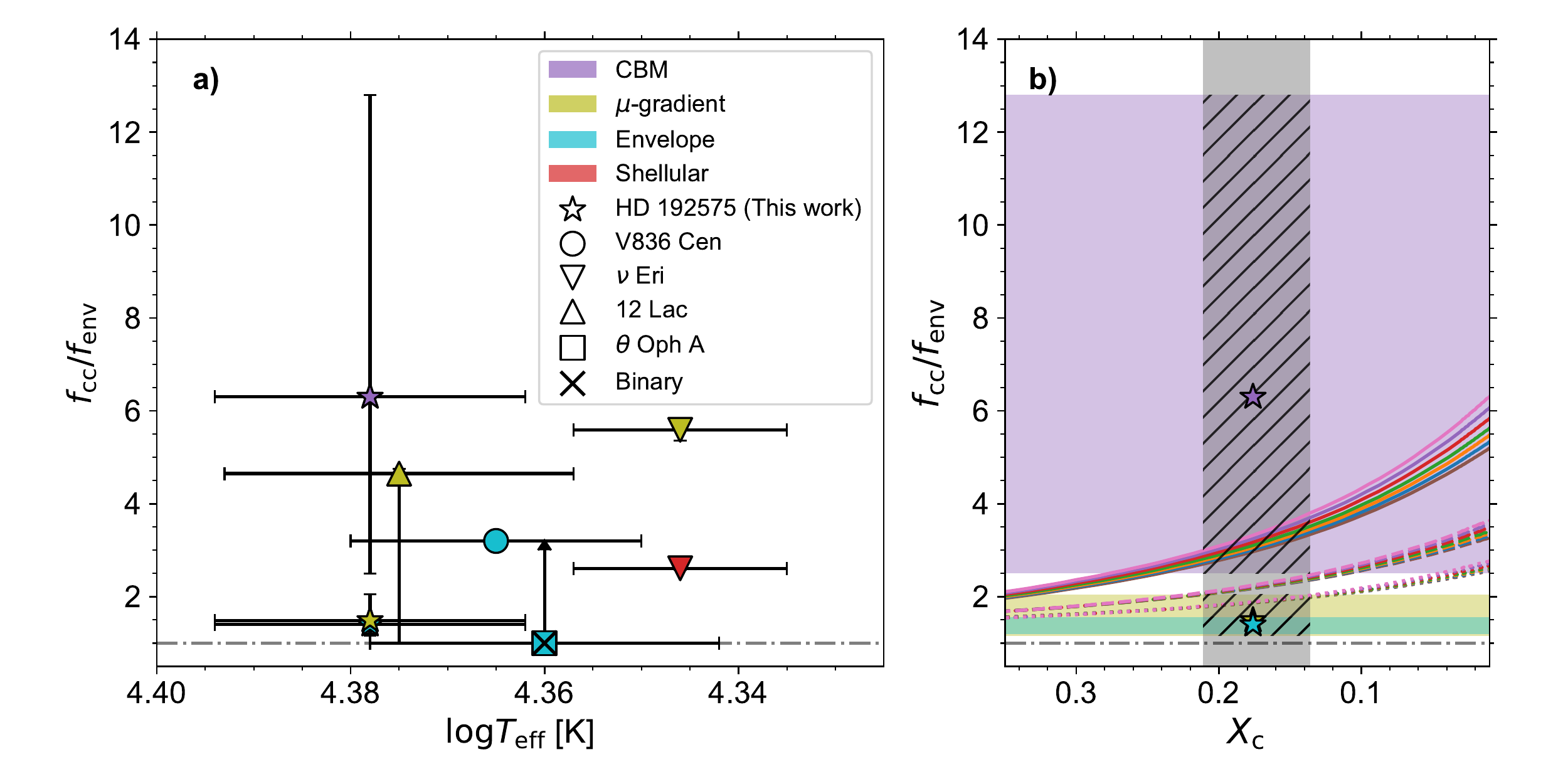}
    \caption{\textbf{Asteroseismic estimates of $f_{\rm cc}/f_{\rm env}$ in the massive star regime, and predictions by rotating stellar evolution models.} Panel (a) shows the $f_{\rm cc}/f_{\rm env}$ value for several hydrogen-core burning stars as a function of the effective temperature. The symbol for each $\beta$~Cep star is colour-coded by the assumed rotational shear layer in the derivation of the rotation profile. The values and confidence intervals used are given in Table~1, and references therein. We note that for the majority of the values in the literature only an upper or lower limit for the core-envelope rotation ratio was available, which we used as the confidence interval limits. The value and confidence interval for HD~192575 was derived from all 558 structure models contained within the derived 2$\sigma$ confidence interval on the stellar parameters. The dashed-dotted line marks rigid rotation ($f_{\rm cc}/f_{\rm env}=1$). Panel (b) shows a comparison against predictions from rotating stellar evolution models which do not include magnetic instabilities \cite{Georgy2013a}. Our results for HD~192575 are shown as star symbols, with the colours corresponding to the selected shear layer, and the uncertainty on $X_{\rm c}$ marked by a grey region. Note that the confidence intervals for the $\mu$-gradient zone and entire radiative envelope  assumptions overlap.  Coloured solid lines show models with $v_{\rm eq, ini}/v_{\rm crit} = 10\%$ with initial masses between 10.5 (brown) and 13.5 M$_{\odot}$ (pink). Dashed and dotted lines show models with $v_{\rm eq, ini}/v_{\rm crit} =30 \%$ and $50\%$, respectively. Several of the models are interpolated between 9, 12, and 15 M$_{\odot}$ tracks in order to conform to the derived mass estimate for HD~192575. Again, a dashed-dotted line marks rigid rotation.}
\end{figure}

\newpage

\normalsize

\section*{Methods}
\label{section: methods}

\subsection*{Prior literature}\label{sec:prior_lit}

HD~192575 is a bright northern early-type star of spectral type B0.5~V\cite{Hohle2010} found at intermediate galactic latitude\cite{GaiaCollaboration2022} ($b=+18.15\degree$), and is located in the TESS northern continuous viewing zone (TESS-NCVZ). We summarise its global observational parameters in Supplementary Table~1. The parallax of HD~192575 was measured in Gaia-DR3, $\pi = 1.56\pm 0.02$, and $d=633\pm7$~parsec \cite{Bailer-Jones2021, GaiaCollaboration2022}. A previous study\cite{Hohle2010} estimated a mass of $12.53\pm0.73$~M$_{\odot}$ using an effective temperature ($T_{\rm eff}\approx 27700$~K) and luminosity ($\log L_{\star}/{\rm L_{\odot}}=4.26$) derived from multi-colour 2MASS and HIPPARCOS photometry. The uncertainty on the mass is relatively small as it is based on purely statistical measurements and does not take into account any systematic uncertainty due to unknown stellar physics which typically dominate the confidence intervals \cite{Serenelli2021}. The pulsational variabilty of HD~192575 was not known prior to our discoveries. We selected HD~192575 as a prime candidate for asteroseismology from a much larger sample of pulsating massive stars observed by TESS.

\subsection*{Spectroscopic analysis}\label{sec:spec_anal}

We obtained two spectra of HD~192575 in July 2019 with the high-resolution HERMES spectrograph ($R\approx85\,000$) mounted on the 1.2-m Mercator Telescope in La Palma, Spain \cite{Raskin2011} as part of a preliminary spectroscopic survey of the TESS Northern-CVZ OB stars. We obtained two additional spectra in September 2020. The observation log is given in Supplementary Table~2. The raw spectra were reduced using the HERMES data reduction pipeline (HERMES-DRS, v.7.0), which includes bias correction, flat-fielding, wavelength calibration, cosmic removal and correction for barycentric motion \cite{Raskin2011}. 

We normalised all spectra manually using spline fits. Radial velocities (RV) were calculated using Lorentzian profiles fit to five He\,I lines (He\,I $\lambda 4009, 4026, 5015, 5876, 6678$~\AA). For each spectral line, the profile of the Lorentzian was fixed through all epochs and all the lines in one epoch are forced to have the same RV shift\cite{Banyard2021}. The derived RVs are given in Supplementary Table~2. They show no significant variations and agree with previous measurements \cite{Wilson1953}. Since the spectra sample several days and then several days a year later we judge short and intermediate period binarity ($P\lesssim1000$~d) unlikely. 

We searched for spectral line-profile variability (LPV) indicating non-radial pulsations \cite{Aerts2003a} and lines in emission indicating the presence of a strong wind \cite{Puls2008}. Early B-type stars with $\beta$~Cep pulsations are known to have LPV with periods of order hours induced by the photospheric velocity fields of (non-radial) pulsations \cite{Aerts2010}. The Si\,III triplet (Si\,III\,$\lambda\lambda 4552, 4567, 4574$~\AA) in particular is commonly used to study the imprint of these pulsations due to their strength in this temperature regime \cite{Aerts2003a}. We show the line profiles of the Si~III triplet of HD~192575 for the different epochs in Supplementary Figure~1. Si\,III $\lambda\lambda 4552$~\AA~in particular shows asymmetric variations, hinting at the presence of non-radial modes. No spectral lines were found to be in emission.

We used the iacob-broad tool\cite{SimonDiaz2014a}, based on a Fourier transform combined with a profile fitting technique, applied to the four HERMES spectra to obtain an estimate of the projected rotational velocity which we found to be $v \sin i \sim 27$~km~s$^{-1}$. This makes HD~192575 a relatively slow rotator. Then, we estimated the effective temperature ($T_{\rm eff}$) and surface gravity ($\log g$) of the star (together with the microturbulence $\xi_{t}$ and surface silicon abundance) by following an analysis strategy based on a combined curve-of-growth analysis of Si\,III-IV lines and a profile fitting of the Balmer lines\cite{SimonDiaz2010a} (see Supplementary Figure~2). To this aim, we employed an extensive grid of NLTE atmosphere models computed with the FASTWIND stellar atmosphere code \cite{Santolaya-Rey1997,Puls2005,RiveroGonzalez2012a, Carneiro2016,Sundqvist2018,Puls2020}. 

We further performed an O and Si abundance analysis for HD~192575 and obtained similar abundances to the B-type stars in the solar neighbourhood (i.e. up to 2 kpc) as investigated by Nieva \& Przybilla (2012)\cite{Nieva2012} and Nieva \& Sim\'{o}n-D\'{i}az (2011)\cite{Nieva2011}. Therefore, while we were not able to compute the Fe abundance with FASTWIND (due to intrinsic limitations of the code), we conclude that this star is similar metallicity and chemical abundance composition to nearby B-type stars. The resultant spectroscopic parameters are given in Supplementary Table~3. The derived $T_{\rm eff}=23900^{+900}_{-900}$~K and $\log g=3.65^{+0.15}_{-0.15}$ of HD~192575 place it within the predicted $\beta$~Cep instability region \cite{Walczak2015}. The final value of the projected rotational velocity is $v\sin i=27^{+6}_{-8}$, confirming HD~192575 as a relatively slow rotator

\subsection*{Distance and luminosity from the Gaia parallax}\label{sec:Gaia}

The position of the star in the Hertzsprung--Russell (HR) diagram is an important constraint needed for asteroseismic modelling. This requires a measurement of $T_{\rm eff}$ which is often supplemented with either the $\log g$ or the star's bolometric luminosity $\log L_{\star}/ {\rm L_{\odot}}$. Given the relatively large uncertainties typical for surface gravities of hot stars, we determined the luminosity of HD~192575 from its available Gaia parallax. We generated a statistical model\cite{Pedersen2020-paper} for the latest Gaia-DR3 passbands ($m_{G}$, $G_{BP}$ and $G_{RP}$) in the temperature regime $T_{\rm eff} \in [10000, 30000]$~K using the Gaia-DR3 filter response function data\cite{Riello2021}. Using the spectroscopic $T_{\rm eff}$ and $\log~g$ values (Supplementary Table~3) we obtained bolometric correction values, $BC_{S_{\lambda}}$, from the statistical model in each Gaia-DR3 passbands (see Supplementary Table~4). The interstellar reddening is obtained from 3D dust maps\cite{Green2019}. The reddening value provided by the 3D dust maps was converted to $E \rm (B-V)$ following \url{http://argonaut.skymaps.info/usage\#units}. We used their Eq.~(2) but the choice of equation ultimately does not affect the calculated luminosity value within its $1\sigma$ confidence interval. We further used the published distance calculated from the Gaia-DR3 parallax \cite{Bailer-Jones2021}. 

Given the parameters in Supplementary Table~4, we calculated a bolometric luminosity of $\log L_{\star}/  {\rm L_{\odot}} = 4.30\pm 0.07$. In the direction of the N-CVZ the interstellar reddening is relatively low compared to the typical reddening in the Galactic bulge, which explains the good precision on the luminosity. Our derived value agrees with the earlier value ($\log L_{\star}/\rm {\rm L_{\odot}}=4.26$) using 2MASS and HIPPARCOS photometry \cite{Hohle2010}. Using the Stefan-Boltzmann law together with the $T_{\rm eff}$ and $\log L_{\star}/{\rm L_{\odot}}$ values we derive an estimate of the stellar radius as $R_{\rm lum, \star} = 8.3\pm 0.9$~R$_{\odot}$.  

For comparison we derive an additional radius constraint using surface-brightness-colour relationships (SBCR) which allow one to easily and independently estimate the limb-darkened angular diameter of the star. Combining the latter with the distance to HD~192575, the linear radius is computed. Considering SBCR relationships dedicated to early-type dwarf stars\cite{Salsi2021}, $m_{V}=6.84\pm 0.009$~mag\cite{Olaj1991}, $m_{K_{s}}=6.639\pm 0.017$~mag\cite{Cutri2003}, and $A_v=0.36$~mag using the \texttt{Stilism} tool\cite{Lallement2014,Capitanio2017}, we find $\theta_{LD}=0.154 \pm 0.005$ mas. Using the Gaia-DR3 distance, i.e. $d=633 \pm 7$~pc\cite{Bailer-Jones2021}, $\pi=1.56 \pm 0.02$ mas~\cite{GaiaCollaboration2022}, we obtain $R=10.5\pm {0.4}\pm{0.1}$~R$_{\odot}$. The first uncertainty comes from the RMS of the SBCR, while the second is related to the uncertainty on the Gaia parallax. The uncertainties on the V and $K_s$ photometry have a negligible impact on the radius (about $1\%$). The discrepancy between this value and the radius from Gaia might be related to the extinction, however the extinction towards HD~192575 is relatively low, and SCBR relationships tend to be only weakly sensitive to extinction (i.e. a 0.2~mag difference in extinction translates to a 2\% difference on the radius). Furthermore, the lack of a clear companion and emission lines, and the relatively low projected surface velocity, eliminate binarity, strong stellar winds and fast rotation as possible sources of the discrepancy. 

\subsection*{Photometric analysis}\label{sec:phot_anal}

HD~192575 was observed by TESS in short-cadence mode (i.e. $2$~min) between sectors 15 and 26, with the exception of sectors 20 and 23. We performed a first inspection of the short cadence TESS data using the \texttt{lightkurve} software package \cite{lightkurve2018} in order to rule out any large aperture mask variations or significant light contamination by other stars. A sector-by-sector analysis showed no significant variations of the aperture mask, and the star is seemingly isolated based on the overlay of Gaia measurements \cite{Gaia2018} (see Supplementary Figure~3). We find a clear central source responsible for the flux, marked in red. The four nearby sources marked in white are notably fainter. The difference in magnitude between these sources and the bright central source is significant enough to disregard them as possible source of contamination in the light curve extraction.

We extracted the cycle~2 TESS 2-min cadence light curve data from the MAST archive (\url{https://archive.stsci.edu/}). We considered only the TESS cycle 2 data, since adding TESS cycle 4 data would introduce a year-long gap. The artefacts in the data analysis resulting from such a gapped 3-year light curve adds unnecessary complexity, while both the frequency resolution and frequency precision deduced from the cycle 2 data alone are largely sufficient for asteroseismic inferences and in any case much smaller than the theoretical uncertainties in the modelling based on Eq.\,(\ref{eq:MD}). We extracted two light curves, one using simple aperture photometry (referred to as SAP), and the pre-conditioned light curve (PDCSAP; Pre-search Data Conditioning Simple Aperture Photometry). The PDCSAP light curve is treated by a pipeline which removes systematics common to all stars on the same CCD \cite{Jenkins2016}. However, thorough checks are needed since the pipeline is optimised for the detection of exoplanets around low mass stars and may not be optimal to study stellar variability. We compare both light curves in Supplementary Figure~4, in which the PDCSAP and SAP light curves are shown for different time scales. HD~192575 exhibits multi-periodic variability on the order of several hours up to several days, unreported in the literature, and is therefore a newly discovered $\beta\,$Cep star. 
The PDCSAP and SAP full cycle 2 light curves spanning 1~yr are similar with the largest differences occurring between Sectors 17-18. Here the SAP light curve shows a slight upward trend towards BJD~2458789, which decreases around BJD~2458792. The TESS pipeline removes this trend which is instrumental in origin. However, the PDCSAP pipeline also systematically removes chunks of data at the end of sectors (e.g. between BJD~2458801 and BJD~2458803). We therefore opted to use the SAP version for the improved duty cycle (75.7\% vs. 68.3\%), and applied our own corrections using low-degree polynomial fits where needed (grey data points in Supplementary Figure~4). This does not affect the short-period pulsations. Our detrended version of the cycle 2 SAP light curve has a time base of $\Delta \rm T =323.77$~d and a frequency resolution $f_{\rm res}=1/\Delta \rm T = 0.0031$~d$^{-1}$.

Our frequency analysis consisted of two stages, extraction of the frequencies, and subsequent filtering by excluding any frequencies not suitable for asteroseismic modelling. In the first stage we used our detrended SAP light curve and extracted the frequencies using \textit{iterative pre-whitening} \cite{Papics2012a, Bowman2021c}. At each step we identified the frequency with the highest signal-to-noise ratio (S/N), using a Lomb-Scargle (LS) periodogram\cite{Lomb1976,Scargle1982} of the light curve. The resulting frequency was used to fit a sine wave model, $\Delta {\rm T_{p}} = A\sin[2\pi f (t-t_{0})+\phi]$, to the time series, from which we obtain amplitude ($A$) and phase ($\phi$) values. This model was subtracted and the residual time-series was used in a next iteration. We repeated the iterations until the S/N of any frequency peak in the LS-periodogram was below five. The S/N was calculated in a window of $1$~d$^{-1}$ around the frequency, thus allowing us to avoid over-interpretation of the TESS data. This significance criterion is based on simulations of space-based time-series data from which the probabilities of false and correct signal identification are measured\cite{Baran2015,Baran2021}.
The frequency extraction yielded a total of 102 frequencies between $0.09$ and $22$~d$^{-1}$.

This frequency list was then checked for unresolved frequencies and alias frequencies (possibly resulting from undersampling of real frequencies\cite{Aerts2010}). This filtering produced 96 well-resolved significant frequencies. The next step involved identifying harmonics and combination frequencies of the form $f_{i} = m \cdot f_{j}+n\cdot f_{k}$ where $f_{i}$ is the combination frequency, $f_{j}, f_{k}$ the `parent' frequencies, and $n, m \in \mathbb{Z}$. These combination frequencies are common in early-type stars with heat-driven pulsations \cite{Papics2012a,Pedersen2021}. We identified a combination or harmonic frequency $f_{i}$ if $|f_{i} - (m \cdot f_{j}+n\cdot f_{k})|< 2.5 f_{\rm res}$, requiring that the amplitude of the combination frequency is less than that of the parent frequencies. In addition we limited ourselves to $n, m \in [\pm 1, \pm 2, \pm 3]$ and only considered the ten highest amplitude frequencies as parent frequencies, in order to avoid chance matches which may occur in dense frequency spectra\cite{VanBeeck2021}. This yielded 12 combination frequencies and 5 harmonics (including two of the dominant frequency). These frequencies were removed as they cannot be used as independent pulsation mode frequencies in our forward asteroseismic modelling methodology.

The resulting frequency list contained 79 independent pulsation mode frequencies.  As a final step we performed a multi-frequency non-linear least-squares fit to the light curve in order to optimise the frequencies, amplitudes and phases, and determine the correlated uncertainties from the resultant covariance matrix which is an important step prior to asteroseismic modelling \cite{Bowman2021c}. The final list of independent pulsation mode frequencies is provided in Supplementary Table~13.
The LS periodogram of our detrended SAP light curve shows significant periodic variability over a relatively large frequency range. We plot parts of the periodogram in Supplementary Figure~5, highlighting the six frequency regions where all measured frequencies occur. A clear dominant frequency is measured at $f_{1}=7.359625(1)$~d$^{-1}$, and four equally spaced frequencies at $f_{3} = 6.350919(2)$~d$^{-1}$, $f_{2}= 6.525005(2)$~d$^{-1}$, $f_{5} = 6.696791(5)$~d$^{-1}$, and $f_{4}=6.867008(4)$~d$^{-1}$ (middle panels of Supplementary Figure~5). Harmonics and combinations of the dominant frequencies are found in the high frequency regime (marked as red and orange triangles, respectively), in addition to several independent modes (lower panels of Supplementary Figure~5).
We also note the presence of low frequencies $1<f<5$~d$^{-1}$ (top panels of Supplementary Figure~5) which fall in the gravity mode frequency domain and one frequency below 1~d$^{-1}$, $f_{15}= 0.09852(2)$~d$^{-1}$. However, this signal only appears in the second half of the light curve (after BJD-2458870), suggesting $f_{15}$ to be of instrumental origin.

The spacing of the equally spaced frequencies ($f_{3},f_{2},f_{5},f_{4}$), $\Delta f \approx 0.17$~d$^{-1}$, motivated us to identify similar spacings in the LS periodogram. We collected all instances in Supplementary Table~5 and sorted them into identified rotational multiplets. This includes the two higher amplitude quadruplets (quad1 and quad2), which are indicative of rotational multiplets of an $\ell\geq2$ mode with one or more missing members. Indeed, the inferred slow rotation from the low $v\sin i=27^{+6}_{-8}$ of HD~192575 indicates that these equally-split multiplets are likely due to rotation. Since $\ell>2$ multiplets tend to be less visible due to photometric cancellation, we assign them as $\ell=2$. This identification is confirmed later using stellar pulsation models. We plot the two high frequency quadruplets in Supplementary Figure~6, showing the notable amplitude difference between the two. Following similar reasoning, we interpret the triplet containing the dominant frequency $f_{1}=7.359625(1)$~d$^{-1}$ as an $\ell=1$ triplet, see Supplementary Figure~7 and Supplementary Table~5.

We also detect a low frequency quintuplet between $f \sim 3.5-4.0$~d$^{-1}$, with a higher average frequency splitting ($\mathcal{S} \sim 0.178-0.195$~d$^{-1}$), as shown in Supplementary Figure~8, with measured frequencies $f_{26}, f_{25},  f_{19}, f_{33}. f_{35}, f_{24},$ and 
$f_{48}$. This multiplet shows two bifurcated 
frequency sets, with small frequency splittings which are of order $\sim 0.01-0.02$~d$^{-1}$. This is the consequence of two overlapping multiplets, quint1a ($f_{26}, f_{19}, f_{35}, f_{24}$) with four members and quint1b ($f_{25}, f_{33}, f_{48}$) with three members. Indeed, if the splittings are a consequence of slow rotation, we expect similar splittings (aside from slight asymmetries) between members of different azimuthal order, $m$. For example, $f_{48}$ is unlikely to be in a multiplet with $f_{26}, f_{19}, f_{35}, f_{24}$ since $f_{48}-f_{24}=0.17855(9)$~d$^{-1}$, while the other splittings are $\Delta f \sim 0.194-0.195$~d$^{-1}$. It is therefore likely part of a closely overlapping second multiplet with $f_{25}, f_{33}$ with at least two missing members, since $(f_{48}-f_{33})/2=0.19545(5)$~d$^{-1}$. We refer to the $f_{26}, f_{19}, f_{35}, f_{24}$ multiplet as quint1a and the $f_{33}, f_{25}, f_{48}$ multiplet as quint1b. By applying similar reasoning as in the case of quad1 and quad2, we assign their angular degree as $\ell=2$. Again this is verified after by comparison with stellar pulsation models. Because we conclude that the quint1a and quint1b multiplets are composed of $\ell=2$ modes, it follows from Eq.~(\ref{eq:split_slow_rot}) that either they have higher $\beta_{nl}$ value compared to the higher frequency multiplets quad1 and quad2, or that the probed mode cavity rotates faster, or both. We test these two hypotheses: faster rotation and/or different $\beta_{\rm nl}$. 

We thus have several candidate rotationally-split multiplets in different frequency regimes, with different average frequency splittings $\mathcal{S} \approx 0.17-0.24$~d$^{-1}$. Next to the average splitting for a multiplet, the asymmetries, $\mathcal{A}$, among the individual rotation splittings are important to consider. In the case of rigid rotation, the second-order rotational splitting equation is given by \cite{Saio1981, Dziembowski1992},
\begin{equation}\label{eq:second_order_split}
f_{m} = f_{0} + \beta_{nl} m f_{\rm rot} + 
D_0 f_{\rm rot}^2 +
D_{L} \frac{m^{2} f_{\rm rot}^2}{f_{0}},
\end{equation}
where 
$D_0$ is caused by the second-order effect of the Coriolis force while $D_{L}$ is a parameter capturing the effect of the centrifugal force on the mode frequency. The second-order rotational effects in Eq.~(\ref{eq:second_order_split}) imply that asymmetries among frequency splittings occur. Detection of such asymmetries therefore reveal second-order effects of rotation and can provide clues for mode identification. Indeed, depending on the value and sign of $D_{L}$ asymmetries should increase or decrease around the zonal ($m=0$) frequency of a multiplet.

We calculate the asymmetries of the multiplets as $\mathcal{A}= f_{m-1}+f_{m+1} - 2 f_{m}$, where the subscript indices refer to the azimuthal order $m$ of a mode with an angular degree $\ell$. We list the calculated $\mathcal{S}$ and $\mathcal{A}$ values for all the multiplets with a least three consecutive members in Supplementary Table~6. For quad1 we find decreasing asymmetries of the order $\mathcal{A}\sim 10^{-3}$~d$^{-1}$ as we go to higher frequencies in the multiplet. If we assume the star is (nearly) rigidly rotating, that $D_{L}$ is the same for each $m$ and that quad1 is $\ell=2$, this implies the zonal frequency of this multiplet is $f_{5}$. A similar reasoning for quad2 implies $f_{8}$ is the zonal frequency in this multiplet. In both cases $D_{L}$ is then negative. Conversely, we find smaller asymmetries of the order $\mathcal{A}\sim 10^{-4}-10^{-5}$~d$^{-1}$, for the triplets trip1, trip2, and trip3. This is as expected from theory for gravity modes \cite{Dziembowski1992}. 

For the multiplet quint1a we find asymmetries of order $\mathcal{A}\sim 10^{-4}$~d$^{-1}$, an order of magnitude lower than for quad1 and quad2. As these lower frequencies correspond to modes with a more dominant gravity-mode character than quad1 and quad2, the smaller asymmetries are understood by the lessened effect of the centrifugal force in the deeper interior of the star. In the case of quint2, we find that $\mathcal{A}_{2}$ differs from $\mathcal{A}_{0}, \mathcal{A}_{1}$ by a sign and an order of magnitude. This implies $f_{37},f_{65}$ may not be part of this multiplet, and that $f_{29},f_{40}, f_{31}$ form a triplet instead. Moreover, the measured asymmetries in quint2 are almost an order of magnitude higher than those measured for the other high frequency multiplets, quad1 and quad2. We therefore find the quint2 multiplet identification uncertain, or the result of combination frequencies, and do not consider it further. We summarise the preliminary mode identification for the detected multiplets in the upper part of Supplementary Table~7 (\textit{``Prior to model comparison''}).

\subsection*{Forward asteroseismic modelling}
\label{sec:forward_modelling}

Forward asteroseismic modelling of pulsation mode frequencies consists of quantitatively comparing observed mode frequencies $f^{\rm obs,i}_{nlm}\pm\sigma^{\rm obs,i}_{nlm}$ to theoretically predicted mode frequencies from stellar models $f^{\rm theo,i}_{nlm}$ for $i=1, ..., N_{f}$, with $N_{f}$ being the number of measured frequencies \cite{Aerts2021}. The fitting procedure is non-trivial as theoretical uncertainties and correlations are difficult to quantify and typically dominate over the accumulated measurement uncertainties \cite{Moya2008,Bowman2021c}. A traditional $\chi^{2}$ merit function does not include the covariance structure in theoretical models and observables, thus may not accurately capture the true overall uncertainty. Hence we use the superior Mahalanobis distance (MD) merit function instead, see Eq.~(\ref{eq:MD}).

Using the multiplets of rotationally-split non-radial pulsation modes (Supplementary Table~7) and the MD as an estimator, we performed forward asteroseismic modelling to constrain a minimal set of important free parameters using identified pulsation modes. These parameters are the mass, core hydrogen content, and the convective boundary mixing, $\bm \theta = (M_{\rm ini}, X_{\rm c}, f_{\rm CBM})$. In a first step we consider a fixed initial chemical composition, but study the effect of a varying chemical composition (through the initial metallicity parameter $Z_{\rm ini}$) on the forward asteroseismic modelling after.

We calculated structure models using the stellar evolution code MESA \cite{Paxton2011,Paxton2013,Paxton2015,Paxton2018,Paxton2019} (v.\,12778). In a first instance we fixed the initial composition ($X_{\rm ini}=0.71$, $Z_{\rm ini}=0.014$, and $Y_{\rm ini} = 1 - X_{\rm ini} - Z_{\rm ini}$) and used the standard set of abundances obtained for B-type stars in the solar neighbourhood\cite{Nieva2012}. This is justified by our spectroscopic results which show little to no difference in O and Si abundances compared to those obtained for typical B-type stars in the solar neighbourhood\cite{Nieva2011, Nieva2012}. We used the OP opacity tables\cite{Seaton2005} and the Nieva \& Przybilla metal mixture \cite{Nieva2012}. For the convective zones the mixing length theory (MLT) formalism was employed including both optically thick and thin material \cite{Henyey1965,Cox1968}, with $\alpha_{\rm MLT}=2.0$. The convective boundaries were determined through the Ledoux criterion and we use the MESA convective premixing scheme (\url{https://docs.mesastar.org/en/latest/reference/controls.html}). 

We also included diffusive exponential convective boundary mixing\cite{Herwig2000}, and applied this to core ($f_{\rm CBM}$), shell and envelope convective zones, setting the $f_{0}$ parameter (which determines how far into the convective zone the mixing is applied) to $f_{0}=0.005$. Since the convective boundary mixing regions around shell and envelope convective zones contain a very small amount of mass, we fixed the value at $f_{\rm shell, env}=2f_{0}$ to include a small, but necessary amount of convective boundary mixing. We implemented envelope mixing in the radiative envelope assuming a profile based on internal gravity waves\cite{Pedersen2018} which originates from hydro-dynamical simulations of a 3.5~M$_{\odot}$ star with $X_{c}=0.5$\cite{Rogers2017}. In this formalism the diffusion coefficient depends on the local density as $D_{\rm mix} (r) \sim D_{\rm env} \cdot \rho^{-1}$. The stitching point is at the base of the envelope $D_{\rm env}$ and can be varied to different levels. More recent hydro-dynamical simulations, covering more masses ($3$, $7$, and $20$~$M_{\odot}$) and different evolutionary stages also support the use of this scaling relation as a reasonable approximation for chemical mixing efficiency set by internal gravity waves in the stellar envelope\cite{Varghese2022}. Choosing a structured internal gravity wave mixing profile over a constant mixing profile for our final grid is motivated by a diffusion coefficient varying throughout the envelope being preferred for 26 SPB stars\cite{Pedersen2021}. We did not rely on structured mixing profiles resulting from stellar structure models including rotation \cite{Paxton2013, Georgy2013a} since these are typically uncalibrated and lead to 
numerical differences in the predicted oscillation frequency values  
well beyond the observational errors
when using a reasonable range of values for the free parameters involved in the diffusion coefficients.

The stellar atmosphere is interpolated from photospheric tables\cite{Paxton2011}. Above $10\,000$~K  the Castelli \& Kurucz model atmospheres \cite{Castelli2003} are used. No rotation is applied to the MESA models following the inferred slow rotation rate of the star and the major uncertainties involved in rotating stellar models. Mass loss was calculated using the Vink~et~al. formalism\cite{Vink2001} scaled by a factor of $0.3$ to conform to recent observational and theoretical results \cite{Puls2015,Bjorklund2021}. Mass loss rates in the considered mass regime and evolutionary stage tend to be very small, for example a $21.5~M_{\odot}$ model with $f_{\rm CBM}=0.02$ in our set-up has a TAMS/ZAMS mass ratio of $\sim 0.990$, while a $9~M_{\odot}$ model with $f_{\rm CBM}=0.02$ in our set-up has a TAMS/ZAMS mass ratio of $\sim0.996$. 

We started by computing a coarse grid covering the core-hydrogen burning phase, from the ZAMS until hydrogen core exhaustion, for masses $M_{\rm ini}$ between 9.0 and 21.5~$M_{\odot}$ in steps of 0.5~$M_{\odot}$ and convective boundary mixing $f_{\rm CBM}$ between 0.005 and 0.035 in steps of 0.005. Mixing at the base of the envelope is fixed at $\log D_{\rm env}=3.00$~dex. We tested different values in the range $\log D_{\rm env}\in[2.00, 5.00]$~dex but found that low order p and g modes with frequency resolutions typical for a 1-yr TESS light curve are not sensitive to these values. We covered core hydrogen fraction intervals of $\Delta X_{c}=0.01$ from $X_{c}=0.701$ to $X_{c}=0.25$, then $\Delta X_{c}=0.005$ to $X_{c}=0.013$. This brings us to a total of 94 stellar models per evolutionary track of a given mass. These choices are motivated by the need to ensure adequate coverage of the spectroscopic parameter space. We show the grid of stellar structure models with fixed initial chemical composition in Supplementary Figure~9. 

Each non-rotating MESA model was perturbed using the stellar oscillation code GYRE\cite{Townsend2013b,Townsend2018,Goldstein2020} (v.6.0.1). Based on previous modelling results of $\beta$~Cep stars, we considered angular degrees $\ell=0-4$, and radial orders $ -7 \leq n_{\rm pg} \leq -1$ for gravity modes, $1 \leq n_{\rm pg} \leq 6$ for pressure modes, with f modes defined as $n_{\rm pg}=0$. We solved the adiabatic oscillation equations, and assume zonal modes ($m=0$). The outer boundary condition is set to that derived by Unno et al.\ (1989)\cite{Unno1989}. The final coarse grid with fixed initial chemical composition contains 182 evolution tracks, 17108 pulsation models and 923832 zonal pulsation frequencies (see Supplementary Table~8). 

The first step in mode identification of HD~192575 was to infer the angular degree, $\ell$, and azimuthal order, $m$, for the multiplets from the observed rotational multiplets in the LS periodogram of the TESS light curve (see Supplementary Table~7). The radial order $n_{\rm pg}$, however, needs to be derived by comparison with predictions of pulsation modes based on the internal structure models. This comparison is further useful to verify our preliminary identification of $\ell$ and $m$ (cf. Supplementary Table~7). We achieved this by radial order fitting the zonal modes of the multiplets one-by-one, using the $2\sigma$ confidence intervals for $T_{\rm eff}$ and $\log L_{\star}/\rm L_{\odot}$ as constraints to select appropriate structure models. We fit the zonal frequency of a rotational multiplet as a given angular degree using the MD and recovered the acceptable radial orders within the $2\sigma$ $T_{\rm eff}$--$\log L_{\star}/\rm L_{\odot}$ confidence intervals. Next we added the zonal frequency of additional multiplets for the same radial orders and determined if it was reconciled. We started with the following multiplets (and their zonal modes) as they have high amplitude, are relatively isolated, and have certain $\ell$ and $m$: quad1, quint1a, quint1b, trip1, quad2 (cf. Supplementary Table~5). 

We show the final mode identification that fits the observed frequencies and satisfies the $2\sigma$ $T_{\rm eff}$--$\log L_{\star}/\rm L_{\odot}$ confidence intervals in Supplementary Table~7. We emphasise that only for this mode identification does the MD merit function yield a large enough sample of models to project robust confidence intervals
that also satisfies the $2\sigma$ $T_{\rm eff}$--$\log L_{\star}/\rm L_{\odot}$ confidence intervals. Thus, our confident mode identification is validated as robust. From this mode identification we subsequently use it to find that $f_{1}$ must be part of an $\ell=1$ multiplet, and have azimuthal order either $m=0$ or $m=+1$. We require higher degrees for quad2 ($\ell>2$) to reproduce the frequency observations in the $2\sigma$ $T_{\rm eff}$--$\log L_{\star}/\rm L_{\odot}$ confidence intervals. Indeed, none of the computed $\ell=2$ modes in this frequency range approach each other closer than $1$~d$^{-1}$ while the distance between the zonal modes of quad1 and quad2 is maximally $f_{5}-f_{8}=0.233381(5)$~d$^{-1}$. In this configuration, we found that only the $(\ell, n_{\rm pg})=(4,1)$ pressure mode can generally reproduce the frequency range of quad2 and the close proximity of quad1 and quad2, with $f_{8}$ or $f_{17}$ as the zonal mode. 

For other, lower amplitude, multiplets the mode identification is less straightforward and we often required higher angular degrees ($\ell=3-4$). This is quite common in slowly rotating $\beta\,$Cep stars\cite{Handler2003, Stankov2005, Cotton2021}. Finally, we also searched for single frequencies that can be explained as radial modes, e.g. the high amplitude single frequencies $f_{8}=8.96669(1)$~d$^{-1}$ and $f_{20}=9.19716(4)$~d$^{-1}$ (cf. Supplementary Figure~5). However, only maximally one of these can be interpreted as a radial mode, since the period ratios of subsequent radial orders do not allow both to be radial for any parameter combination within the $2\sigma$ observational confidence intervals in the HR~diagram and the previously identified rotational multiplets. Therefore, we conclude that at this point we reached the limit of a robust and confident identification of radial orders of rotational multiplets given the currently available constraints. We then proceeded to fit all confidently identified rotational multiplets given their mode identification in Supplementary Table~7. 

Armed with the confidently identified multiplets, we determined the overall best model in the case of a fixed initial chemical composition and the projected confidence intervals for each model parameter. First we excluded models that do not satisfy the avoided crossing criterion, see Eq.~(\ref{eq:avoided_crossing}) (e.g., models outside the grey region in Figure~2). This left 2852 remaining structure models.

We then calculated the MD values for these remaining models in the grid (Eq.~\ref{eq:MD}). We used frequencies $f_{5} (\ell=2, n_{\rm pg}=2), f_{19} (\ell=2, n_{\rm pg}=-2)$ as the identified zonal frequencies in the quad1 and quint1a multiplet, respectively. For the quint1b multiplet $(\ell=2, n_{\rm pg}=-2)$ we defined the zonal frequency as $(f_{48}+f_{33})/2 = 4.01414$~d$^{-1}$ and used it for the calculation. We refer to these modes as $ f_{\rm obs, p2}, f_{\rm obs, g2}$ and $f_{\rm obs, g1}$, respectively, {and used these three to calculate the MD values. We did not fit quad2 as it was difficult to confidently determine the zonal mode despite almost all models predicting it be close to $f_{5}$. After the calculation of the MD values for the 2852 models, we required that the remaining models satisfy the $2\sigma$ observational constraints of $\log T_{\rm eff}$ and $\log L_{\star}/{\rm L_{\odot}}$ in the HR~diagram, leaving 235 models. The effective temperature $\log T_{\rm eff}$ and stellar luminosity $\log L_{\star}/{\rm L_{\odot}}$ were therefore not directly used in the calculation of the MD values.

To determine parameter uncertainties, we used the likelihood function associated with the MD (Eq.~\ref{eq:MD}) in combination with Bayes' theorem \cite{Mombarg2021,Michielsen2021}. That is, we found the likelihood of parameters $\bm{\theta} = (\theta^{1}, \theta^{2}, ..., \theta^{k})$ given the observed data $\bm{D}$:
\begin{equation}
    \mathcal{L}(\bm{\theta| D}) = \exp \Big( -\frac{1}{2} (\ln(|V+\Lambda|) + k \ln(2\pi) + \rm MD) \Big),
\end{equation}
where $k$ is the number of free parameters, and the MD (Eq.~\ref{eq:MD}) of a set of parameters $\bm \theta$ \cite{Aerts2018b,Mombarg2021,Michielsen2021}. As this is proportional to the probability density function $\mathcal{L}(\bm{\theta| D}) \sim P(\bm{D| \theta})$ we applied Bayes' theorem allowing us to describe the probability of a parameter $\theta^{m}$ occurring in range $ \theta^{m} \in [\theta^{m}_{a}, \theta^{m}_{b}$] as:
\begin{equation}
    P(\theta^{m}_{a} < \theta^{m} < \theta^{m}_{b}|\bm{D}) = \frac{\Sigma^{q}_{i} P(\bm{D|\theta_{i}}) P(\bm{\theta_{i}})}{\Sigma^{Q}_{j} P(\bm{D|\theta_{j}}) P(\bm{\theta_{j}})},
\end{equation}
where the denominator is a summation over the models $j = 1,\ldots,Q$ that are consistent with the $2\sigma$ confidence interval on the spectroscopic parameters and the avoided crossing criterion (Eq.~\ref{eq:avoided_crossing}), and the numerator is a summation over all models $i=1,\ldots,q$ with the highest likelihood such that $P(\theta^{m}_{a} < \theta^{m} < \theta^{m}_{b}|\bm{D})=0.95$ (equivalent to a $2\sigma$ confidence interval on the derived parameters). In our case $\bm{\theta} = (M_{\rm ini}, X_{\rm c}, f_{\rm CBM})$ and $k=3$. 

We provide the best-fitting model's parameters and $2\sigma$ confidence intervals in the second column of Supplementary Table~9. Note that the derived mass is the initial mass, but that due to the low mass-loss rate in this mass range, the actual mass is within the same confidence interval. The parameter correlations of the models that satisfy the avoided crossing criterion combined with the $2\sigma$ confidence intervals on $T_{\rm eff}$ and $\log L_{\star}/{\rm L_{\odot}}$ is shown in Supplementary Figure~10, where the correlation between $M_{\rm ini}$ and $f_{\rm CBM}$ is clearly seen. Among these models, lower mass models generally have higher convective boundary mixing values, and vice versa. Because of this correlation and since it drives stellar evolution, we infer a convective core mass of $M_{\rm cc}=2.8^{+0.5}_{-0.7}$~M$_{\odot}$. This value was obtained by taking all stellar structure models within the derived $2\sigma$ confidence interval and calculating the mass contained within the convective core as defined by the Ledoux criterion. We plot the histogram of this inferred core mass distribution within the confidence intervals of $M_{\rm ini}, f_{\rm CBM}$ and $X_{\rm c}$ in Supplementary Figure~11 (in case of a fixed initial composition). This distribution is narrower than the distribution of $f_{\rm CBM}$ and similar to the distribution of $M_{\rm ini}$ (cf. Supplementary Figure~10). The inferred core mass implies a core mass fraction of $M_{\rm cc}/M_{\rm \star}= 23 \pm 6~\%$ which conforms to estimates from binary star studies in the same mass regime \cite{Tkachenko2020,Johnston2021}. The seismic radius $R_{\star, \rm seism}$ and age of the star, $t_{\star}$ were inferred similarly by retrieving these values from all stellar structure models within the $2\sigma$ confidence interval. The seismic radius agrees with the luminosity radius derived using the Gaia distance but does show a discrepancy with the surface brightness radius.

We did not use multiplets trip1 and quad2 in the best model determination because of their tentative mode identification. We show the theoretical frequencies of the best model (in case of a fixed initial composition) in the frequency range of trip1 and quad2 in Supplementary Figure~12. We find that trip1 can be explained as an $(\ell, n_{\rm pg}) = (1,4)$ mode, with the dominant mode $f_{1}$ having azimuthal order $m=0$ or $m=+1$. We require higher degrees for quad2, $(\ell, n_{\rm pg})=(4,1)$ to reproduce the frequency observations in the $2\sigma$ $T_{\rm eff}$--$\log L_{\star}/\rm L_{\odot}$ confidence intervals, as alluded to before. These are general finding for trip1 and quad2 in all models that reproduce the fitted multiplets. In the best model the zonal mode ($m=0$) of quad2 is either $f_{8}$ or $f_{17}$.

We further verified the robustness of derived parameters and their confidence intervals against possible mode misclassification. This included changing the assigned $m=0$ modes in the fitted multiplets, including/excluding $f_{1}$ as an $\ell=1$ mode, and fitting quad2 instead of quad1 as the $\ell=2$ pressure mode. While the best models shifted slightly between configurations, the confidence intervals did not change significantly from the values quoted in Supplementary Table~9.

In the preceding analysis we used the standard chemical mixture as derived by Nieva \& Przybilla (2012) \cite{Nieva2012} based on the similarity between our derived O and Si abundances, and the abundances found for typical B-type stars in the solar vicinity\cite{Nieva2011, Nieva2012}. However, there is a known mass-metallicity(-overshooting) correlation in the analysis of $\beta$~Cep stars\cite{Ausseloos2004}, leading to over/underestimated masses if the metallicity is under/overestimated, respectively. Moreover, the current chemical composition of HD~192575 (as measured on the surface) may be different from its initial composition. Finally, a recent study on the modelling of $\beta$~Cep stars has shown that in order to accurately describe the central chemically homogeneous regions, in our case by inferring $M_{\rm cc}$ and $R_{\rm cc}$, one needs to ensure the chemical composition of the models is representative of the observed chemical composition\cite{Salmon2022a}. 

To explore the effect of the chemical composition and associated theoretical uncertainties, we computed additional models with initial metallicities in the confidence interval given by Nieva \& Przybilla (2012)\cite{Nieva2012} (their Table~10), $Z_{\rm ini} \in [0.012, 0.016]$ in steps of $0.002$ for $M_{\rm ini}\in[9.0, 17.5]$ and $f_{\rm CBM}\in[0.005, 0.035]$. The initial helium mass fraction was determined through the Galactic enrichment law, $Y_{\rm ini}=Y_{\rm p} + \frac{{\rm d}Y}{{\rm d}Z} Z_{\rm ini}$, where $Y_{\rm p}$ and $\frac{{\rm d}Y}{{\rm d}Z}$ are the primordial helium abundance and enrichment ratio, respectively. We assumed $Y_{\rm p}=0.2453$\cite{Aver2021} and required that the enrichment ratio was able to reproduce the mass fractions of the adopted chemical mixture $(X_{\rm ini}, Y_{\rm ini}, Z_{\rm ini}) = (0.710, 0.276, 0.014)$\cite{Nieva2012}. Solving for $\frac{{\rm d}Y}{{\rm d}Z}$ then yielded $\frac{{\rm d}Y}{{\rm d}Z}=2.193$ which is at the higher end of the confidence interval derived recently using asteroseismology\cite{Verma2019}. This yielded a grid of 39312 models. 

We then repeated the forward modelling, using the same robustly identified modes, now considering $Z_{\rm ini}$ as an additional parameter such that $\theta = (Z_{\rm ini}, M_{\rm ini}, X_{\rm c}, f_{\rm CBM})$. Applying the avoided crossing criterion, and the $2\sigma$ observational constraints on $T_{\rm eff}$ and $\log L_{\star}/{\rm L_{\odot}}$ in succession led to 4962 and 663 remaining structure models, respectively. We show the derived and inferred parameters in Supplementary Table~9, and a parameter correlation plot in Figure~3. A final total of 558 stellar structure models are contained within the $2\sigma$ confidence intervals of the parameters in Supplementary Table~9. 

While the best model has slightly different parameters, the confidence intervals on the free parameters remain the same, demonstrating the robust nature of our results. We find a slightly lower mass and metallicity, but a larger convective boundary mixing value. This result is a direct manifestation of the known mass-metallicity-overshoot relation\cite{Ausseloos2004}. The confidence interval on $Z_{\rm ini}$ covers the whole considered parameter range, suggesting our methodology is not sensitive to such a parameter when using only three identified pulsation frequencies, but this is to be expected. This is also confirmed in Figure 3, showing that models with different $Z_{\rm ini}$ do not show much variation in their MD value. Importantly, the additional varying of $Z_{\rm ini}$ around the typical value for B-type stars in the solar neighbourhood does have a slight effect on the confidence intervals of the inferred parameters, notably on the inferred core and stellar radii. In both cases, the confidence intervals slightly increased. Nevertheless, the value and confidence interval of the inferred core mass remain the same demonstrating it to be robust ($M_{\rm cc}/M_{\rm \star}= 23 \pm 7~\%$). As in the fixed metallicity case we find that the unused multiplet trip1 can be explained in the best fitting models as an $(\ell, n_{\rm pg}) = (1,4)$ mode, with the dominant mode $f_{1}$ having azimuthal order $m=0$ or $m=+1$. Again, we require higher angular degrees for quad2, $(\ell, n_{\rm pg})=(4,1)$ to reproduce the observed frequencies of these modes.

The best structure model and parameter confidence intervals resulted from use of the avoided crossing (Eq.~\ref{eq:avoided_crossing}), constraints from the $T_{\rm eff}$ and $\log L_{\star}/{\rm L_{\odot}}$ parameters, and the modelling of three frequencies using a statistical framework taking the theoretical uncertainties of the stellar models into account. We note that the well-constrained $X_{\rm c}$ value is a direct result of the detection of modes near an avoided crossing. Indeed, it is much less sensitive to the input parameters of our modelling setup than the inferred parameters $M_{\rm cc}, R_{\rm cc}, R_{\star, \rm seism}$ and $t_{\star}$. We show this by means of an example in Supplementary Figure~13. In this figure we show the stellar structure models that remain after applying only the avoided crossing criterion (cf. Eq.~\ref{eq:avoided_crossing}). This leads to two ranges in $X_{\rm c}$, around $X_{\rm c}\sim 0.175$ and $X_{\rm c}<0.10$. If we consider the first range, panel a) in Supplementary Figure~13 shows that $X_{\rm c}$ must have a value around $X_{\rm c}\sim 0.17$, with little sensitivity to initial mass, or convective boundary mixing value (coefficient of variation $\sigma_{X_{\rm c}}/\mu_{X_{\rm c}}=0.11$). This is different for the inferred convective core mass, as shown in panel b) of Supplementary Figure~13, which is clearly more sensitive to the input parameters (coefficient of variation $\sigma_{M_{\rm cc}}/\mu_{M_{\rm cc}}=0.23$). 

The power of an avoided crossing in constraining the stellar $X_{\rm c}$ value was similarly demonstrated in the case of $\beta$~CMa ($M_{\star}=13.5\pm0.5$~M$_{\odot}$) \cite{Mazumdar2006}. For this star, a small difference between the $\ell=0$ and the $\ell=2$ modes was measured which could only occur directly at an $\ell=2$ avoided crossing, leading to a particularly well-constrained $X_{\rm c}=0.126\pm0.003$. Our methodology is similar to theirs. Yet, their increased precision compared to our value for HD~192575 is due to the fact that we have two $\ell=2$ modes either approaching an $\ell=2$ avoided crossing, or just passing an $\ell=2$ avoided crossing (cf. Figure~2), while in the case of $\beta$~CMa\cite{Mazumdar2006} the difference between the $\ell=0$ and $\ell=2$ modes can only be explained directly at the $\ell=2$ avoided crossing (see their Figure~5).

\subsection*{Constraints on non-rigid rotation profile}\label{sec:rotation}

In a first instance, using the best-fitting structure model with $Z_{\rm ini}$ as a varied parameter (cf. Supplementary Table~9) we derived the rotation frequency directly from the three confidently identified frequency multiplets using the model-dependent Ledoux constants, in the case of rigid rotation, such that $f_{\rm rot} (x) = f_{\rm rigid}$ for all $x$, where $x$ is the normalised radius. In this case the frequency splittings in the first-order perturbative approach (Eq.~\ref{eq:split_slow_rot}) are given by 
\begin{equation}\label{eqINTRO:splitting}
f_{nlm} = f_{ nl0} + \Big(m \beta_{nl} f_{\rm rigid} \Big).
\end{equation} 
For each multiplet consisting of $N$ non-zonal frequencies we then calculate the non-zonal frequencies $f_{n l m}$, the observed $f_{ nl0}$, the $\beta_{\rm nl}$ constant from the best model, and a range of rotation frequencies $f_{\rm rigid} \in [f_{\rm rigid, low}, f_{\rm rigid, high}]$. We then calculate the reduced $\chi^{2}$ value for each $f_{\rm rigid}$ value by summing over the $N$ non-zonal frequencies as
\begin{equation}
    \chi^{2}_{\rm red} = \frac{1}{N_{\rm obs} - k}\sum_{i=1}^{N} \left(\frac{f_{\rm obs}^{i} - f_{\rm theo}^{i}}{f_{\rm res}}\right)^{2},
\end{equation}
where $N_{\rm obs}$ is the number of non-zonal modes, $k=1$ the number of free parameters to estimate ($f_{\rm rigid}$), $f_{\rm obs}^{i}$ the observed non-zonal frequency, $f_{\rm theo}^{i}$ the calculated non-zonal frequency, and $f_{\rm res}=0.00308$~d$^{-1}$ the frequency resolution of the light curve. We show the result of this exercise for the three confidently identified multiplets in the upper panels of Supplementary Figure~14. The stellar model is the best model derived including variation of $Z_{\rm ini}$ (Supplementary Table~9).
We considered a range $f_{\rm rigid}\in [0.10, 0.30]$~d$^{-1}$ in 300 steps. A clear minimum was reached with a $3\sigma$ confidence interval. This confidence interval is defined as
\begin{equation}
    \chi^{2}_{3\sigma} = \frac{\chi^{2}_{3\sigma, w} \cdot \chi^{2}_{\rm red, min}}{w},
\end{equation}
where $\chi^{2}_{\rm red,min}$ is the $\chi^{2}$ value of the best fit, $w=N_{\rm obs}-1$ is the degrees of freedom and $\chi^{2}_{3\sigma, w}$ is the tabulated value of the $3\sigma$ confidence interval of the cumulative distribution function of a $\chi^{2}$ distribution with $w$ degrees of freedom. For the three multiplets, each with four members, $w = 3-1=2$, and $\chi^{2}_{3\sigma, 2}=11.83$. By calculating the intersection of $\chi^{2}_{3\sigma}$ with a fit through the $\chi^{2}$ distribution we obtain a confidence interval on the derived rotation frequency. Using the radius from the best structure model we can also derive the equatorial velocity $v_{\rm  eq}= 2\pi f_{\rm rigid} R_{\rm model}$. The lower panels of Supplementary Figure~14 show the theoretical frequencies obtained for the optimised rotation frequencies

The numerical results of this exercise for all confidently identified multiplets are shown in Supplementary Table~10. From these results we conclude that the observations are inconsistent with rigid rotation within their confidence intervals. We emphasise that irrespective of the structure model used, a systematically higher rotation frequency is always derived for the gravity-mode multiplets than for the pressure-mode multiplet.

To quantify the differential radial rotation profile of HD~192575 we used an analytical three-zone model of the rotation rate $f_{\rm rot}(x)$ as a function of the fractional radius $x$, see Eq.~(\ref{eq:omega_model}). This model assumes a uniform rotation rate in the core $f_{\rm cc}$, a linear dependence of $f_{\rm rot}(x)$ on the fractional radius in the considered rotational shear layer, and a uniform envelope rotation rate on top of the considered rotational shear layer, $f_{\rm env}$. We consider three assumptions for the definition of the rotational shear layer:
\begin{enumerate}
\item The convective boundary mixing zone on top of the convective core,
\item The $\mu$-gradient zone on top of the convective core \cite{Pamyatnykh2004,Dziembowski2008},
\item The entire radiative envelope on top of the convective core \cite{Aerts2003b, Dupret2004}.
\end{enumerate}
The last option represents a linear decrease from the convective core throughout the star up to the surface. In this assumption the derived $f_{\rm env}$ is the rotation frequency at the outermost radial point of the model. Implementing these options in the three zone model (Eq.~\ref{eq:omega_model}) then comes down to changing the extent of $x_{\rm shear}$. We choose to test the sensitivity of our results to these three scenarios, as such a comparison has never been performed in the literature before.

We measured the unknown $f_{\rm cc}$ and $f_{\rm env}$ values from the measured frequency splittings within three confidently identified rotational multiplets and Eqns.~(\ref{eq:split_slow_rot}) and (\ref{eq:omega_model}), which yielded the following linear system of equations for measured frequency splittings $\Delta f_{\rm p2}, \Delta f_{\rm g1}, \Delta f_{\rm g1}$:
\begin{eqnarray}\label{eq:three_zone_system}
\Delta f_{\rm p2} = f_{\rm cc}\cdot K_{0, \rm p2} + f_{\rm env} K_{1, \rm p2} \, 
\\
\Delta f_{\rm g1} = f_{\rm cc}\cdot K_{0,\rm g1} + f_{\rm env} K_{1,\rm g1},
\\
\Delta f_{\rm g2} = f_{\rm cc}\cdot K_{0,\rm g2} + f_{\rm env} K_{1,\rm g2},
\end{eqnarray}
where $K_{0, \rm p/g}, K_{1, \rm p/g}$ are integrals over the fractional radius:
\begin{eqnarray}\label{eq:K_system}
K_{0, \rm p/g} = \int^{x_{\rm shear}}_{0} \beta_{\rm nl}K_{\rm nl}(x) {\rm d}x + \int^{x_{\rm shear}}_{x_{\rm cc}} \frac{x - x_{\rm cc}}{x_{\rm cc}-x_{\rm shear}} \beta_{\rm nl}K_{\rm nl}(x) {\rm d}x \\
K_{1, \rm p/g} = \int^{R_{\star}}_{x_{\rm shear}} \beta_{\rm nl}K_{\rm nl}(x) {\rm d}x - \int^{x_{\rm shear}}_{x_{\rm cc}} \frac{x - x_{\rm cc}}{x_{\rm cc}-x_{\rm shear}} \beta_{\rm nl}K_{\rm nl}(x) {\rm d}x,
\end{eqnarray}
which are calculated from the rotational kernel profiles once $x_{\rm shear}$ is defined (cf. right panel of Figure~2 and Figure~4). This forms a linear matrix equation of form $Ax=b$, where $x,b$ are vectors and $A$ is a matrix. In our case the matrix $A$ consists of the $K$ values shown in Eq.~(\ref{eq:three_zone_system}):
\begin{equation}
    \begin{pmatrix}
K_{0, \rm p2} & K_{1, \rm p2}\\
K_{0, \rm g1} & K_{1, \rm g1}\\
K_{0, \rm g2} & K_{1, \rm g2}\\
\end{pmatrix}
\cdot 
\begin{pmatrix}
f_{\rm cc}\\
f_{\rm env}\\
\end{pmatrix}
= 
   \begin{pmatrix}
 \Delta f_{\rm p2}\\
 \Delta f_{\rm g1}\\
 \Delta f_{\rm g2}\\
\end{pmatrix},
\end{equation}
whose individual values are given by Eq.~(\ref{eq:K_system}). Note that the system is overdetermined (i.e. we have more rows in $A$ than unknowns). Indeed, technically we only need two multiplets of modes probing sufficiently separate pulsation cavities to solve the system. Nevertheless, including additional multiplets of modes allows us to maximally exploit the available information. Moreover, in the case of HD~192575 the g$_{1}$ and g$_{2}$ ($\ell=2$) modes are either about to interact and exchange mode character, or have just done so (cf. Figure~2). Indeed, the g$_{1}$ multiplet of modes has a mixed character (p and g) before the avoided crossing, but a dominant g mode character after the interaction (and vice versa for the g$_{2}$ mode, see right panel in Figure~2). This means that in our structure models the multiplet of modes with the dominant g mode character is either the g$_{1}$ or the g$_{2}$ mode depending on the models evolutionary state with respect to the occurrence of an avoided crossing. For the reasons mentioned above, we consider both g modes in the calculation.

The measured frequency splittings $\Delta f_{\rm p2}, \Delta f_{\rm g1}, \Delta f_{\rm g1}$ were calculated by averaging the small differences in the frequencies between each azimuthal order. This yielded $\Delta f_{\rm p1} = 0.172030(3)$~d$^{-1}$ for the p$_{2}(\ell=2$) multiplet, $\Delta f_{\rm g1} = 0.19490(3)$~d$^{-1}$ for the g$_{1}(\ell=2$) multiplet, and $\Delta f_{\rm g2} = 0.19516(4)$~d$^{-1}$ for the g$_{2}(\ell=2$) multiplet.  

We then calculated the $K$ values (Eq.~\ref{eq:K_system}) and solved the linear system in Eq.~(\ref{eq:three_zone_system}) for the three assumed rotational shear layers for all 558 structure models contained within the 2$\sigma$ confidence interval (cf. Supplementary Table~9) of the best structure model (with $Z_{\rm ini}$ as a varied parameter). Since the system is overdetermined there is no `exact' solution, and we therefore used a least-squares method from the \texttt{numpy} python package (\texttt{numpy.linalg.lstsq}\cite{Numpy2020}) to solve the system for each stellar structure model.

Contrary to previous studies who were limited to a handful or only a single best model, we calculate the full range of rotation rates among all statistically significant structure models. We find a large distribution in $f_{\rm cc}/f_{\rm env}$ values among the 558 structure models in the case of a convective-boundary mixing rotational shear layer, as shown in the top panel of Supplementary Figure~15. We obtain a skewed distribution with a lower boundary of $f_{\rm cc}/f_{\rm env}=2.58$ and a higher boundary of $f_{\rm cc}/f_{\rm env}=12.88$ within the 2$\sigma$ confidence interval. The distribution shows that most models are between the lower boundary and $f_{\rm cc}/f_{\rm env}\sim 7$, with a long tail of higher values.

In order to understand the resulting distribution we considered the condition number of a system of linear equations $Ax=b$, where $x,b$ are vectors and $A$ is a matrix, which is given by:
\begin{equation}
\kappa(A) = \frac{|\lambda_{\rm max}(A)|}{|\lambda_{\rm min}(A)|}.
\end{equation}
The values $\lambda_{\rm max}$ and $\lambda_{\rm min}$ are the maximum and minimum eigenvalues of matrix $A$, respectively. Generally, the condition number, $\log \kappa$, provides an estimate of the number of digits of accuracy lost in addition to the loss due to arithmetic methods. In our case, the condition number is clearly correlated with the gravity mode $K_{0}$ values in Eq.~(\ref{eq:K_system}) as shown in the lower left panel of Supplementary Figure~15. This is expected as the $K_{0}$ indicates how much probing power a certain mode has in the convective core and rotational shear layer. This in turn is determined by the mode character, with pure pressure modes having low $K_{0}$ values, while modes with dominant gravity mode character have relatively higher $K_{0}$ values, as seen in Supplementary Figure~15. We note that lower condition numbers are obtained for systems where the g$_{1}$ and g$_{2}$ mode are of sufficiently different $K_{0}$ values (i.e. mode character). That is, the condition number increases in the case of structure models where the g modes are interacting to a higher degree (nearer to or at the avoided crossing), which leads to overall lower probing power of the gravity modes in the near-core region. We show the $\log \kappa$ of each model versus the derived $f_{\rm cc}/f_{\rm env}$ in the lower right panel of Supplementary Figure~15. This shows that models with high $f_{\rm cc}/f_{\rm env}$ values tend to have high condition numbers because they have less probing power in the convective core. Therefore, the values with a higher condition number and concurrently larger values of $f_{\rm cc}/f_{\rm env}$ are actually less precise measurements of $f_{\rm cc}/f_{\rm env}$. 

Nevertheless, there is another effect of the avoided crossing which must be considered. Structure models where the g$_{1}$ mode has a more dominant gravity-mode character (i.e. after the avoided crossing, shown as orange circles in the lower right panel of Fig.~15) generally yield lower $f_{\rm cc}/f_{\rm env}$ values. On the other hand, structure models where the g$_{2}$ has a more dominant gravity-mode character (i.e. before the avoided crossing, shown as blue crosses in the lower left panel of Supplementary Figure~15) generally yield higher $f_{\rm cc}/f_{\rm env}$ values. As a result we cannot discard all models with $f_{\rm cc}/f_{\rm env}\sim5-10$ as being solutions of ill-conditioned linear systems. We therefore define the confidence intervals by taking the median as our central value and taking the minimal/maximal $f_{\rm cc}/f_{\rm env}$ values as our lower/upper confidence values. We take the same approach in defining the confidence intervals on the separate $f_{\rm cc}$ and $f_{\rm env}$ values.

Next we considered the $\mu$-gradient zone as the rotational shear layer by setting $x_{\rm shear}$ to the fractional radius corresponding to the convective core mass at ZAMS\cite{Pamyatnykh2004,Dziembowski2008}. The results are shown in Supplementary Figure~16. We find a lower boundary of $f_{\rm cc}/f_{\rm env}=1.15$ and a higher boundary of $f_{\rm cc}/f_{\rm env}=2.05$ for the 2$\sigma$ structure model confidence interval. This range is narrower, resulting in a precise measurement of the rotation rates, which we show in Supplementary Table~11. This is a result of the order-of-magnitude lower condition numbers of the linear systems ($\log \kappa \in [0.5,1.0]$ versus $\log \kappa \in [1.6,2.2]$) due to the higher $K_{0}$ values relative to the convective boundary mixing zone assumption. This in turn is a consequence of the rotational kernels having their dominant probing power in the $\mu$-gradient zone (cf.~Figure~4). While the effect of the structure models' evolutionary state with respect to the avoided crossing is still clear, as seen in the lower right panel of Supplementary Figure~16, it leads to a less broad range of rotation frequencies. As a result, the relative $f_{\rm cc}/f_{\rm env}$ rates are lower  to the case where we assumed the convective-boundary mixing region as the rotational shear layer, but rigid rotation is still firmly excluded.

Finally we considered the entire radiative envelope as the rotational shear layer by setting $x_{\rm shear}=R_{\star, \rm model}$\cite{Aerts2003b, Dupret2004}. The results are shown in Supplementary Figure~17. We find a lower boundary of $f_{\rm cc}/f_{\rm env}=1.2$ and a higher boundary of $f_{\rm cc}/f_{\rm env}=1.56$ for the 2$\sigma$ structure model confidence interval. The derived rotation rates are shown in the last row of Supplementary Table~11. Like in the case of the $\mu$-gradient zone assumption we find lower relative $f_{\rm cc}/f_{\rm env}$ rates, but confidently exclude rigid rotation. The effect of the avoided crossing is again seen, but less pronounced than in the two other assumptions on the rotation shear layer.

We summarise the results of the three-zone model with three different assumptions on the extent of the rotational shear layer in Figure~4 (d). This demonstrates that the assumption of the rotational shear layer has a major impact on the derived asteroseismic rotation rates, while the uncertainties propagating from the $2\sigma$ confidence intervals of the best structure models are much smaller in comparison. We emphasise that previous studies have treated this problem in the literature using only a single assumption on the rotational shear layer (cf. Table~1). Our results quantify and compare the impact of the mutually exclusive assumptions for the location of the shear layer for a given $\beta$~Cephei star. All these rotation profiles firmly exclude rigid rotation in HD~192575.

\subsection*{Inclination of the stellar rotation axis}

If we assume that HD~192575 is only differentially rotating in the radial direction, we can use the derived $f_{\rm env}$ and the inferred stellar radius to estimate the equatorial velocity $v_{\rm eq}=2\pi f_{\rm env}R_{\star}$ and inclination:
\begin{equation}
    i = \arcsin \left[ \frac{v\sin i}{v_{\rm eq}}\right],
\end{equation}
where $v\sin i$ is the measured projected rotational velocity. We use the stellar radius, $R_{\star} = R_{\star, \rm seism}=9.1^{+0.8}_{-1.7}$~R$_{\odot}$ inferred in the forward asteroseismic modelling, and the $v\sin i = 27^{+6}_{-8}$~km~s$^{-1}$ derived from spectroscopy}. The results are summarised in Supplementary Table~12. They show that HD~192575 is seen at a relatively low inclination with $i=17(6)\degree$ or $19(6)\degree$, depending on the assumed shear layer.

We point out that the inclination angle of a pulsating star can also be deduced from
the ratio of mode heights between azimuthal orders within a rotational multiplet of a certain angular degree and radial order, in the case of equipartition of energy within a multiplet\cite{Gizon2003}.
While this method is readily applied to solar-like pulsators whose modes are driven by convection\cite{Gehan2021},
it is invalid for the heat-driven modes of hot massive pulsators. This is demonstrated by the observed amplitudes within the multiplet components with equal $|m|$ for HD\,192575, as well as for other $\beta$\,Cep stars in the literature\cite{Aerts2004a,Handler2004,Handler2005}. We are therefore unable to evaluate the derivation of the inclination angle from this independent method.

\newpage

\section*{Supplementary Material} 

Supplementary material is available for this paper.

Table S1 – S13\\
Figure S1 – S17

\section*{Data availability}\label{sup:inlists}
The cycle 2 TESS data for HD~192575 can be retrieved from the MAST archive (\url{https://archive.stsci.edu/}). For information regarding the HERMES spectra we refer to \url{http://www.mercator.iac.es}. The full frequency list in machine-readable format and the MESA/GYRE inlists needed to reproduce our results and figures are available through the open repository Zenodo (doi:10.5281/zenodo.7823538).


\section*{Code availability}
The iterative pre-whitening code is freely available and documented at \url{https://github.com/IvS-KULeuven/IvSPythonRepository}. Information about access to the FASTWIND stellar atmosphere code can be found at \url{https://fys.kuleuven.be/ster/research-projects/equation-folder/codes-folder/fastwind}. The \texttt{iacob-broad} tool from the IACOB project is freely available from: \url{http://research.iac.es/proyecto/iacob/pages/en/useful-tools.php}. The stellar evolution code, \texttt{MESA}, is freely available and documented at \url{http://mesa.sourceforge.net/}, and the stellar pulsation code, \texttt{GYRE}, is freely available from \url{https://github.com/rhdtownsend/gyre} and documented at \url{https://gyre.readthedocs.io/en/stable/index.html}.

\section*{Acknowledgements}
The research leading to these results has received funding from the European Research Council (ERC) under the European Union's Horizon 2020 research and innovation program (grant agreement No. 670519: MAMSIE). DMB gratefully acknowledges a senior postdoctoral fellowship from the Research Foundation Flanders (FWO) with grant agreement no. 1286521N. MM gratefully acknowledges a PhD scholarship from the Research Foundation Flanders (FWO) under project No. 11F7120N. The research leading to these results has (partially) received funding from the KU Leuven Research Council (grant C16/18/005: PARADISE). SS-D acknowledges support from the Spanish Government Ministerio de Ciencia e Innovaci\'on through grants PGC-2018-091\,3741-B-C22, PID2021-122397NB-C21, and from the Canarian Agency for Research, Innovation and Information Society (ACIISI), of the Canary Islands Government, and the European Regional Development Fund (ERDF), under grant with reference ProID2020010016. VV gratefully acknowledges support from the Research Foundation Flanders (FWO) under grant agreement N°1156923N. RHDT acknowledges support from NSF grant ACI-1663696 and NASA grant 80NSSC20K0515. GH acknowledges financial support by the Polish NCN grants 2015/18/A/ST9/00578 and 2021/43/B/ST9/02972. JSGM gratefully acknowledges funding from the French Agence Nationale de la Recherche (ANR), under grant MASSIF (ANR-21-CE31-0018-02). The MESA and GYRE developers are thanked for their efforts in providing, maintaining, and supporting the use of the open-source stellar-evolution code and pulsation codes. This research has made use of the SIMBAD database, operated at CDS, Strasbourg, France. The TESS data presented in this paper were obtained from the Mikulski Archive for Space Telescopes (MAST) at the Space Telescope Science Institute (STScI), which is operated by the Association of Universities for Research in Astronomy, Inc., under NASA contract NAS5-26555. Support to MAST for these data is provided by the NASA Office of Space Science via grant NAG5-7584 and by other grants and contracts. Funding for the TESS mission is provided by the NASA Explorer Program. Based on observations made with the Mercator Telescope, operated on the island of La Palma by the Flemish Community, at the Spanish Observatorio del Roque de los Muchachos of the Instituto de Astrof\'{i}sica de Canarias. Based on observations obtained with the HERMES spectrograph, which is supported by the Research Foundation - Flanders (FWO), Belgium, the Research Council of KU Leuven, Belgium, the Fonds National de la Recherche Scientifique (F.R.S.-FNRS), Belgium, the Royal Observatory of Belgium, the Observatoire de Genève, Switzerland and the Thüringer Landessternwarte Tautenburg, Germany. This work presents results from the European Space Agency (ESA) space mission Gaia. Gaia data are being processed by the Gaia Data Processing and Analysis Consortium (DPAC). Funding for the DPAC is provided by national institutions, in particular the institutions participating in the Gaia MultiLateral Agreement (MLA). The computational resources and services used in this work were provided by the VSC (Flemish Supercomputer Center), funded by the Research Foundation - Flanders (FWO) and the Flemish Government – department EWI.

\section*{Author Contributions}
SB discovered variability of HD~192575 in TESS mission data, performed the photometric analysis, computed the model grid, performed the asteroseismic modelling, and wrote the manuscript. DMB defined the project, supervised SB, contributed to the photometric and seismic analysis, and guided the interpretation. MM and RHDT contributed to the modelling setup and seismic analysis. SSD performed the spectroscopic analysis and contributed the final atmospheric parameters. CA provided context, guided the exploitation of the avoided crossings, and helped with the interpretation. VV contributed to the derivation of the stellar rotation profiles. GB contributed to the gathering of spectra and the RV analysis. NN provided the radius measurement from surface-brightness-colour relationships. GH aided in the frequency analysis and the identification of the rotationally-split multiplets. JSGM contributed to the gathering of spectra. RV and GR are the Deputy PI and PI of the TESS mission, respectively. All authors discussed and commented on the manuscript.

\section*{Competing interests}
The authors declare no competing interests.

\end{document}